\documentclass[journal=jacsat,manuscript=article]{achemso}
\usepackage[version=3]{mhchem} 
\usepackage{subfig}
\usepackage{graphicx}
\usepackage{dcolumn}
\usepackage{bm}
\usepackage[table]{xcolor}

\usepackage{natbib,hyperref}
\hypersetup{
  colorlinks = true,
  linkbordercolor = {white},
  breaklinks=true,
  linkcolor = {black},
  anchorcolor = {black},
  citecolor = {black},
  filecolor = {black},
  menucolor = {black},
  runcolor = {black},
  urlcolor = {blue}
}
\usepackage{multicol}
\newcommand{\degree}{^\circ}
\usepackage{longtable}

\usepackage[T1]{fontenc} 
\usepackage{lmodern}
\usepackage[utf8]{inputenc}

\title[]
  {Computational Investigation of Copper Phosphides as Conversion Anodes for Lithium-Ion Batteries}

\author{Angela F. Harper}
  \email{afh41@cam.ac.uk}
   \affiliation{Theory of Condensed Matter, Cavendish Laboratory, University of Cambridge, J. J. Thomson Avenue, Cambridge CB3 0HE, U.K.}
 \author{Matthew L. Evans}
  \email{me388@cam.ac.uk}
   \affiliation{Theory of Condensed Matter, Cavendish Laboratory, University of Cambridge, J. J. Thomson Avenue, Cambridge CB3 0HE, U.K.}
\author{Andrew J. Morris}%
   \email{a.j.morris.1@bham.ac.uk}
   \affiliation{School of Metallurgy and Materials, University of Birmingham, Edgbaston, Birmingham B15 2TT, U.K}

\begin{document}

\begin{abstract}

Using first principles structure searching with density-functional theory (DFT) we identify a novel $Fm\bar{3}m$ phase of Cu$_2$P and two low-lying metastable structures, an $I\bar{4}3d$--Cu$_3$P phase, and a $Cm$--Cu$_3$P$_{11}$ phase. The computed pair distribution function of the novel $Cm$--Cu$_3$P$_{11}$ phase shows its structural similarity to the experimentally identified $Cm$--Cu$_2$P$_7$ phase. The relative stability of all Cu--P phases at finite temperatures is determined by calculating the Gibbs free energy using vibrational effects from phonon modes at 0\,K. From this, a finite-temperature convex hull is created, on which $Fm\bar{3}m$--Cu$_2$P is dynamically stable and the Cu$_{3-x}$P ($x < 1$) defect phase $Cmc2_1$--Cu$_8$P$_3$ remains metastable (within 20\,meV/atom of the convex hull) across a temperature range from 0\,K to 600\,K. Both CuP$_2$ and Cu$_3$P exhibit theoretical gravimetric capacities higher than contemporary graphite anodes for Li-ion batteries; the predicted Cu$_2$P phase has a theoretical gravimetric capacity of 508\,mAh/g as a Li-ion battery electrode, greater than both Cu$_3$P (363\,mAh/g) and graphite (372\,mAh/g). Cu$_2$P is also predicted to be both non-magnetic and metallic, which should promote efficient electron transfer in the anode. Cu$_2$P's favorable properties as a metallic, high-capacity material suggest its use as a future conversion anode for Li-ion batteries; with a volume expansion of 99\,\% during complete cycling, Cu$_2$P anodes could be more durable than other conversion anodes in the Cu--P system with volume expansions greater than 150\,\%. The structures and figures presented in this paper, and the code used to generate them, can be interactively explored online using \href{https://mybinder.org/v2/gh/harpaf13/data.copper-phosphides/master?filepath=CuP_results.ipynb}{Binder}.

\end{abstract}

\maketitle

\section{Introduction}
\label{intro}

Graphite is the most commonly employed lithium-ion battery (LIB) anode, but is inherently limited by a maximum theoretical capacity of 372\,mAh/g upon formation of LiC$_6$. Phosphorus (black or red) has a significantly higher theoretical capacity of 2596\,mAh/g due to the formation of Li$_3$P; however it suffers from capacity deterioration, primarily caused by deleterious volume expansion that occurs upon charging, which constrains the capacity to 350--500\,mAh/g in a limited voltage window \cite{Ramireddy2015}. In addition, P and its lithiated phases have limited electrical conductivity, requiring dopants and additives to improve performance. By adding transition metals to P, through nanostructuring or synthesis, both electrical conductivity and stability during cycling can be enhanced \cite{BHATT201910852}. 

Transition metal phosphides (TMP) provide a large design space in which to engineer such high-capacity, conversion anodes for LIBs \cite{tarascon2011issues}. High-throughput computational screening has previously identified TMPs with high capacities for LIB electrodes including TiP, Co$_2$P, Mn$_2$P and others \cite{Wolverton2013}. As conversion anodes for LIBs, TMPs offer both added gravimetric capacity (ranging from 500 to 2000\,mAh/g \cite{Wolverton2013}) and stability against volume expansion over several battery cycles \cite{Sun2016}. In addition to bulk or powdered TMPs being used as LIB conversion anodes \cite{FengAdvances2017}, nanostructured TMPs can often display improved electrochemical cycling performance \cite{LU2018972}. Despite these efforts, TMPs have yet to be widely adopted as conversion anodes, given the large volume expansion (between 150\,\% to 300\,\% \cite{Wolverton2013}) exhibited by anodes with high P content, which limits their cyclability. Despite this drawback in volume expansion, TMPs show higher average voltages than graphite, which has an average voltage of 0.1\,V. For example, CoP has an average voltage of 0.67\,V, the ternary metal phosphide LiFeP has an average voltage of 0.4\,V, and MnP has an average voltage of 0.62\,V \cite{Wolverton2013}. Higher average voltages give the metal phosphides improved safety while sacrificing energy density, making them an ideal choice for large-scale and long-term energy storage.

Several previously studied TMP anodes include FeP$_{x=1,2,4}$~\cite{Boyanov2006}, Fe$_2$P nanoparticles \cite{Zhang2015}, Ni$_2$P~\cite{Lu2012}, CuP$_2$ \cite{Kim2017CuP2,WANG2003480}, and Cu$_3$P \cite{BICHAT200480,PFEIFFER2004263}, among others. Of the TMPs tested as conversion anodes, the copper phosphides (specifically CuP$_2$ and Cu$_3$P) have shown promise for their cyclability and capacity. The copper phosphides offer additional benefits to the other TMPs, as Cu is already used as a common current collector, providing further cycling capability and resistance to degradation \cite{villevieille2008good}. Cu$_3$P prepared by high temperature synthesis had a first-cycle capacity of 527\,mAh/g \cite{BICHAT200480}, and a porous Cu$_3$P anode synthesized by facile chemical corrosion exhibited a capacity between 360--380\,mAh/g over 70 cycles \cite{Ni2014Cu3P}. The capacity of high temperature synthesized Cu$_3$P exceeds that of graphite, and the cyclability of porous Cu$_3$P is improved relative to other Cu$_3$P anodes \cite{BICHAT200480,chandrasekar2013thin}. CuP$_2$ on the other hand, delivers a higher initial capacity of 815\,mAh/g, but can only be cycled stably 10 times before the capacity fades to 360\,mAh/g\cite{WANG2003480}. The main factor in this degradation is the high concentration of P in the CuP$_2$ which, while enabling high capacity, also contributes to the structural instability of CuP$_2$ during cycling as the lithium-rich Li$_3$P phase forms. To optimize the trade-off between stability and capacity, it would be beneficial to discover a compound with higher P content than Cu$_3$P to offer higher capacity, and with a Cu content higher than CuP$_2$ to aid in cyclability.

By performing crystal structure prediction, combining both \textit{ab initio} random structure searching (AIRSS) and a genetic algorithm (GA), in addition to structural prototyping with known crystal structures of related chemistries \cite{CURTAROLO2012218,saal2013materials,hautier2011data}, we produced the compositional phase diagram of the copper phosphide system. We describe this approach to structure prediction and the application of open source Python packages \texttt{matador} (v0.9) \cite{matador}, for high-throughput first principles calculations, and \texttt{ilustrado} (v0.3) \cite{ilustrado}, for computational structure prediction with GAs. Crystal structure prediction for battery anodes is a well-tested method \cite{harper2019ab}, used for identifying both novel anode materials \cite{Wolverton2013}, and unknown phases which form during battery cycling \cite{see2014ab,mayo2017structure}. AIRSS has been used previously to search for additional phases of Li--P and Na--P which form during battery cycling \cite{mayo2016ab}. The GA was also employed to search for new phases of Na--P, which were confirmed experimentally through solid-state nuclear magnetic resonance (NMR) spectroscopy \cite{MarbellaACS2018}. As applied here to Cu--P, these methods predict a novel metallic $Fm\bar{3}m$--Cu$_2$P phase at 0\,K, within the target composition range of Cu$_{1<x<3}$P, for a high-capacity, low volume expansion conversion anode; we compare its electronic structure to other TMPs to show a similarity to $Fm\bar{3}m$-Rh$_2$P and $Fm\bar{3}m$-Ir$_2$P. Two other phases, $Cm$--Cu$_3$P$_{11}$ and $I\bar{4}3d$--Cu$_3$P are identified as metastable, both bearing structural similarity to known copper phosphides. We calculate the convex hull of Cu--P at temperatures up to 600\,K, confirming the dynamic and chemical stability of Cu$_2$P across this temperature range. A ground-state voltage profile from density-functional theory (DFT), shows that $Fm\bar{3}m$--Cu$_2$P undergoes the same lithiation process as $P6_3cm$--Cu$_3$P; however $Fm\bar{3}m$--Cu$_2$P has a higher capacity of 508\,mAh/g, with an average voltage of 0.86\,V versus Li/Li$^+$ (compared to 0.91\,V for $P6_3cm$--Cu$_3$P).

\section{Methods}
\label{methods}

To search for novel copper phosphides, we first performed structural relaxations of the 13 structures from the Inorganic Crystal Structure Database (ICSD) \cite{hellenbrandt2004inorganic} of Cu$_x$P ($0 < x < 1$). The Python package \texttt{matador} \cite{matador} was used to query 1053 prototype binary structures from the Open Quantum Materials Database (OQMD) \cite{saal2013materials} with chemical compositions containing a pnictogen and a transition metal from the first two rows, namely \{Ti, V, Cr, Mn, Fe, Co, Ni, Cu, Zn, Zr, Nb, Mo, Tc, Ru, Rh, Pd, Ag, Cd\}--\{P, As, Sb\}; each composition was then transmuted to the corresponding stoichiometry of Cu--P, yielding 909 unique structures after geometry optimization. In order to extend this search beyond existing prototypes, two additional structure prediction steps were performed, namely AIRSS \cite{pickard2011ab} and an evolutionary search with the GA implemented in the \texttt{ilustrado}\cite{ilustrado} package.

When performing AIRSS, one proceeds by generating random ``sensible'' (symmetry, density and atomic separation constrained) trial cells and then geometry optimizing them to their corresponding local minima. All relaxations can be performed concurrently, with no interdependence between calculations. New trial structures are generated until the ground state of each stoichiometry (within the constraints of the search) has been found multiple times.

We initially performed an exploratory AIRSS search consisting of around 5000 trial structures, with constraints on cell size, stoichiometry, and number of atoms in the cell. In this initial search, the total number of atoms in the cell was constrained to be $\leq$ 40, and the total number of formula units was randomized between 1 and 4, while still keeping the total number of atoms below 40.  The number of atoms of Cu and P were randomized between 1 and 9 in each cell, and the cell volume ($V$) was constrained based on the total number of atoms in the cell ($N$) to be $8N\,\AA^3 \leq V \leq 20N\,\AA^3$, based on the average densities of Cu--P phases within the ICSD.

Structures from the searching and enumeration procedures were then used, with fitness weighted according to their distance from the convex hull, as the initial configurations for a GA implemented in the Python package \texttt{ilustrado} \cite{ilustrado}. The \texttt{ilustrado} package uses a simple cut-and-splice crossover operation, supplemented by mutation operators (random noise, atomic permutations, vacancies and adatoms) \cite{Deaven1995}. To avoid stagnation, each trial structure was filtered for similarity (via pair distribution function overlap) against existing structures in the population. Three independent GA runs were performed with 10 generations each, yielding a further 1049 relaxed structures. Finally, a directed AIRSS search of Cu$_x$P$_y$ where $x+y<8$, was performed to create a final set of \textasciitilde20,000 structures within the Cu--P chemical space. In all cases, to constrain the search to physically reasonable structures, a minimum atomic separation of 1.5\,\AA\,was enforced and the maximum number of atoms in the cell was constrained to 10 for the initial \textasciitilde10,000 AIRSS searches and 40 atoms per cell for the final \textasciitilde3,000 trials.

All calculations were performed using CASTEP (v18.1 and v19.1), the plane wave pseudopotential DFT package \cite{clark2005first}. To maximize computational efficiency, the initial calculations were performed with loose convergence criteria that ensured formation energies converged to 10\,meV/atom. The Perdew-Burke-Ernzerhof (PBE) exchange-correlation functional was used \cite{perdew1996generalized} with Vanderbilt ultrasoft pseudopotentials \cite{vanderbilt1990soft} that required a plane wave kinetic energy cutoff of 300\,eV to converge energies to within 10\,meV/atom. The Brillouin zone (BZ) was sampled with a Monkhorst-Pack grid $k$-point spacing finer than $2\pi\times0.05$\,\AA$^{-1}$; the grid was frequently recomputed to accommodate any changes in cell shape and size during relaxation. Each structure was geometry optimized at this accuracy to a force tolerance of 0.05\,eV/\AA. The structures with a formation energy within 50\,meV of the convex hull were then further optimized once more using CASTEP's on-the-fly (OTF) ``C18'' library of ultrasoft pseudopotentials\footnote{OTF pseduopotential strings are Cu: 3|2.2|2.0|1.0|10|12|13|40:41:32(qc=6), P: 3|1.8|4|4|5|30:31:32, Li: 1|1.0|14|16|18|10U:20(qc=7)} with a finer $k$-point sampling of 2$\pi\times0.03$\,\AA$^{-1}$ and plane wave kinetic energy cutoff of 500\,eV, which yielded formation energies converged to within 2.5\,meV/atom. In order to predict the voltage profiles with the same convergence criteria (formation energies within 2.5\,meV/atom), the relaxation of known Li--P structures required a higher plane wave cutoff of 700\,eV. Therefore, to compare ternary phases of Cu--Li--P in the voltage profile, all Cu--Li--P phases were re-optimized at a plane wave kinetic energy cutoff of 700\,eV.

To identify stable structures from this search, a convex hull of the copper phosphides was created. The formation energy $E_f$ of each structure Cu$_x$P$_y$ was calculated using,

\begin{equation} \label{eq:Ef}
E_f(\mathrm{Cu}_x\mathrm{P}_y) = E(\mathrm{Cu}_x\mathrm{P}_y) - xE(\mathrm{Cu}) - yE(\mathrm{P}),%
\end{equation}
where $E($Cu$)$ is the DFT total energy of the $Fm\bar{3}m$--Cu structure, and $E($P$)$ is the energy of $Cmca$-P (black phosphorus). Black phosphorus was used as the P chemical potential instead of the lower energy polymorph red phosphorus; as has been previously discussed in Mayo \textit{et al.} \cite{mayo2016ab}, black phosphorus is commonly used when making electrochemical cells \cite{sun2012electrochemical}.

Electrochemical voltage profiles for Li insertion into the stable Cu--P phases were calculated from the computed formation energies from ternary convex hull of Cu--Li--P. In order to calculate the voltage profiles shown in the section on Cu$_2$P as a Li-ion battery conversion anode, the  voltage, $\bar{V}$, between two tie-lines in the ternary convex hull with compositions Li$_{x_1}$Cu$_n$P and Li$_{x_2}$Cu$_n$P was calculated using,

\begin{equation} \label{eq:voltage}
\bar{V}(x_1,x_2) = -\frac{E(\mathrm{Li}_{x_1}\mathrm{Cu}_n\mathrm{P}) - E(\mathrm{Li}_{x_2}\mathrm{Cu}_n\mathrm{P}) - (x_1 - x_2)E(\mathrm{Li}) }{(x_1-x_2)F},
\end{equation}
as stated by Urban et al. \cite{urban2016computational}. In equation \ref{eq:voltage}, $E(\mathrm{Li}_{x_1}\mathrm{Cu}_n\mathrm{P})$ and $E(\mathrm{Li}_{x_2}\mathrm{Cu}_n\mathrm{P})$ are the ground state energies of two phases on along the reaction pathway of the ternary convex hull, in which $x_1$ and $x_2$ are the relative amounts of Li in the starting and ending products at each point in the pathway.

All phonon calculations were performed under the harmonic approximation with the PBE $xc$-functional in a $2\times2\times2$ supercell (corresponding to a phonon $q$-point spacing of 2$\pi$\,$\times$\,0.046\,\AA$^{-1}$ for $Fm\bar{3}m$-Cu$_2$P) using the finite displacement method implemented in the CASTEP code. The dynamical matrix was then Fourier interpolated onto the BZ path provided by the SeeK-path Python package \cite{HINUMA2017140,togo_spglib_2018} to compute the phonon dispersion, and onto a fine Monkhorst-Pack grid to compute the phonon density of states. 

The band structure for Cu$_2$P was calculated using the higher accuracy parameters and pseudopotentials mentioned previously, and the electronic density of states was integrated and projected onto atomic orbitals using the OptaDOS code \cite{morris2014optados,NichollsOptaDOS}. Vibrational properties of all stable phases were computed using the finite displacement method, with an added many-body dispersion correction (MBD denoted MBD* in CASTEP v19.0) \cite{tkatchenko2012accurate} to account for inter-layer interactions in black phosphorus.

The open source Python package \texttt{matador} (v0.9) \cite{matador} was used to run the CASTEP calculations, perform the analysis and create the plots found in this article. All of this analysis, as well as the underlying source code and data, can be explored interactively using \href{https://mybinder.org/v2/gh/harpaf13/data.copper-phosphides/master?filepath=CuP_results.ipynb}{Binder} and found on Github \href{https://github.com/harpaf13/data.copper-phosphides}{\texttt{harpaf13/data.copper-phosphides}}. The input and output files associated with our calculations have been deposited into the Cambridge Repository at \href{https://doi.org/10.17863/CAM.52272}{https://doi.org/10.17863/CAM.52272}.

\section{Results}
\label{results0K}

From the search of \textasciitilde20,000 trial structures, there are 42 unique phases within 50\,meV/atom of the convex hull. Previous computational structure searches have used a distance above the hull of 25\,meV/atom \cite{sun2016thermodynamic}, and given the accuracy of PBE \cite{zhang2018performance}, we chose to increase this cutoff to 50\,meV/atom. Furthermore the experimentally verified $P6_3cm$--Cu$_3$P structure\cite{olofsson1972crystal} is 37\,meV/atom above the convex hull tie-line, further justifying this cut-off. Uniqueness was determined by computing pairwise overlap integrals of the pair distribution functions of phases at each stoichiometry using \texttt{matador}. The set of 42 unique phases contains four experimentally reported copper phosphides from the ICSD; $\text{P}\bar{1}$--CuP$_{10}$ synthesized by a mineralization reaction \cite{CuP10}, $\text{C}2/m$--Cu$_2$P$_7$ \cite{mouller1982darstellung}, $\text{P}2_1/c$--CuP$_2$ \cite{mouller1982darstellung} and $\text{P}6_3cm$--Cu$_3$P \cite{SchlengerCu3P,olofsson1972crystal} from high temperature sintering. 

\begin{figure}
    \centering
    \includegraphics[scale=1]{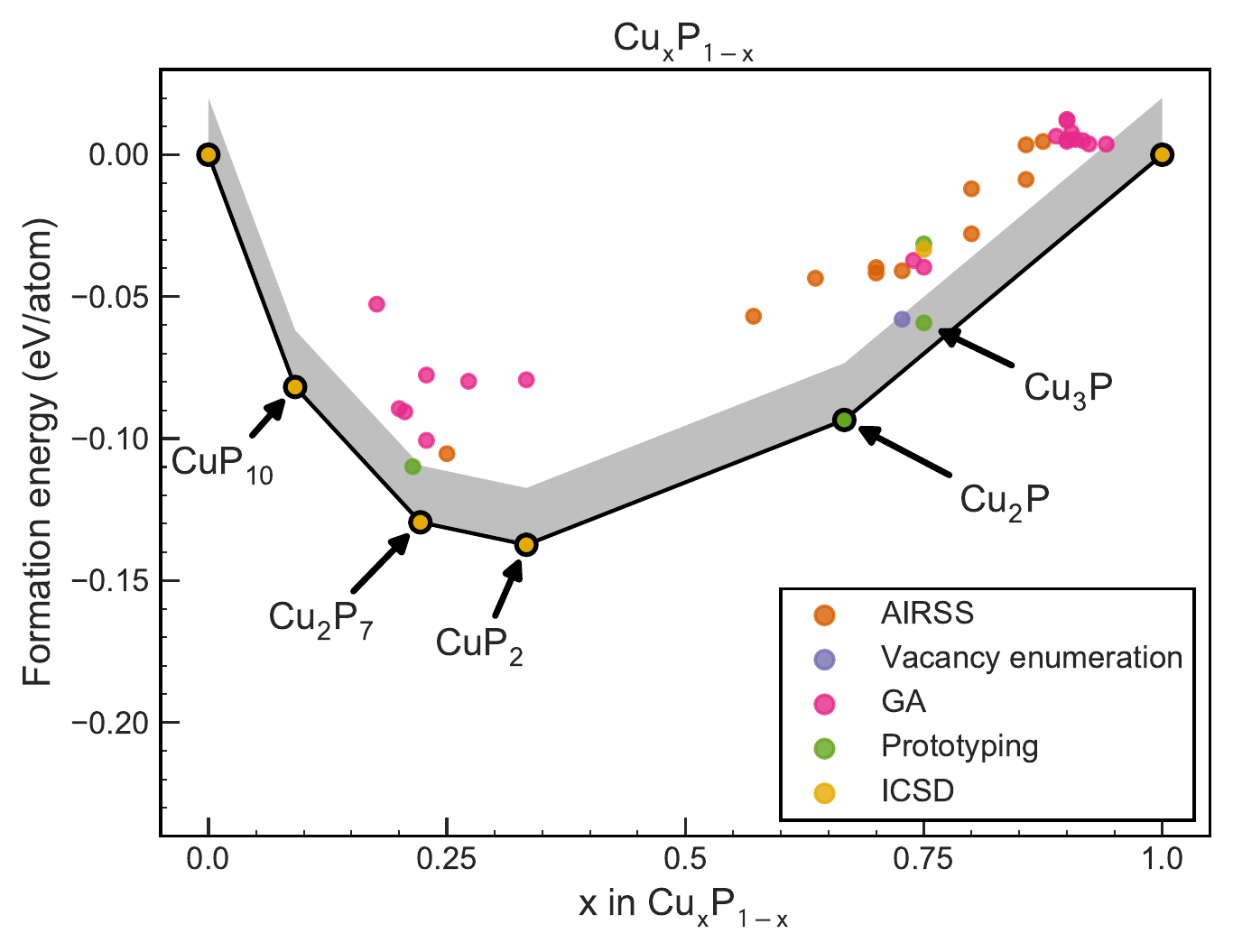}
    \caption{Convex hull of Cu--P phases from structure searching. Four structures lie on the convex hull, CuP$_{10}$, Cu$_2$P$_7$, CuP$_2$, and Cu$_2$P. Structures are colored according to their provenance: either from a searching method (AIRSS, GA, Prototyping, Vacancy enumeration) or from an existing database (ICSD). Prototyping refers to using a prototype structure from the ICSD, and replacing the atoms with Cu or P as described in the Methods section. The Vacancy Enumeration phases are the phases optimized after adding Cu vacancies to $P6_3cm$--Cu$_3$P. Phases within 20\,meV/atom of the convex hull lie within the shaded gray region, and Cu$_3$P is labeled for reference.}
    \label{fig:CuP-convex-hull}
  \end{figure}

Oloffson's experiments on single crystal $P6_3cm$--Cu$_3$P synthesized at high temperature, and subsequent work by deTrizio et al. \cite{DeTrizio2015}, show that Cu$_3$P has several defects \cite{olofsson1972crystal} with a range of stoichiometries between Cu$_{2.6}$P and Cu$_{2.8}$P. DFT studies of the Cu vacancies indicate that Cu$_3$P is substoichiometric \cite{DeTrizio2015} and to search this substoichiometric space, unit cells of $P6_3cm$--Cu$_{18}$P$_6$ were enumerated with 1, 2, and 3 Cu vacancies, resulting in 76 Cu$_{3-x}$P structures. The lowest energy defect was a $Cmc2_1$--Cu$_8$P$_3$ (Cu$_{2.67}$P) phase 26\,meV/atom above the convex hull tie-line, denoted as Vacancy enumeration in Figure \ref{fig:CuP-convex-hull}.

The convex hull of Cu--P, with points colored by the provenance of each structure, is presented in Figure \ref{fig:CuP-convex-hull}; the experimentally identified phases, and a new $Fm\bar{3}m$--Cu$_2$P phase, all lie on the convex hull tie-line and are each labeled with an arrow.

 \begin{table}[!htb]
    \centering

  \begin{minipage}{20cm}

\begin{tabular}{l c c c c }
\footnotetext{Light gray indicates experimentally confirmed phases} 
\footnotetext{Dark gray indicates new phases with $\Delta E \leq$ 20\,meV/atom} \\
\rowcolor{gray!20}
Formula & On  & $\Delta E$ & Spacegroup & Provenance\\
\rowcolor{gray!20}
& Tie-line & (meV/atom) &             &            \\
& & & & \\
\rowcolor{gray!20}              Cu         &       $\star$   &          -           &       $Fm\bar{3}m$         &           ICSD 43493\footnote{Experimental lattice parameter for Cu from \cite{otte1961lattice}}      \\
        Cu$_\text{6}$P      &              &          31          &        $I4/mmm$        &             AIRSS              \\
        Cu$_\text{4}$P      &              &          28          &        $P4/nmm$        &             AIRSS              \\
        Cu$_\text{4}$P      &              &          44          &         $Cmmm$         &             AIRSS              \\
\rowcolor{gray!50}        Cu$_\text{3}$P      &              &          11          &        $I\bar{4}3d$         &          Prototype 64715\footnote{Prototype structure is $I\bar{4}3d$--Cu$_3$As\cite{steenberg1938crystal} }           \\
        Cu$_\text{3}$P       &             &          30          &        $P2_1/m$        &               GA               \\
\rowcolor{gray!20}        Cu$_\text{3}$P     &               &          37          &        $P6_3cm$        &           ICSD 15056\footnote{Structure from single crystal diffractometry \cite{olofsson1972crystal}}           \\
        Cu$_\text{3}$P            &        &          39          &         $I\bar{4}$          &          Prototype 23560\footnote{Prototype structure $I\bar{4}$--Cr$_3$P by single crystal X-ray diffraction\cite{Owusu_1972}}           \\
  Cu$_\text{17}$P$_\text{6}$    &          &          40          &          $P1$          &               GA               \\
  Cu$_\text{8}$P$_\text{3}$   &            &          26          &        $Cmc2_1$        &             AIRSS              \\
  Cu$_\text{8}$P$_\text{3}$    &           &          36          &        $P6_3cm$        &             AIRSS              \\
  Cu$_\text{8}$P$_\text{3}$    &           &          39          &          $P1$          &             AIRSS              \\
  Cu$_\text{7}$P$_\text{3}$    &           &          42          &          $P1$          &             AIRSS              \\
  Cu$_\text{7}$P$_\text{3}$    &           &          44          &          $P1$          &             AIRSS              \\
\rowcolor{gray!50}        Cu$_\text{2}$P        &  $\star$   &          -           &       $Fm\bar{3}m$         &          Prototype 38356\footnote{Prototype structure $Fm\bar{3}m$--Rh$_2$P by X-ray diffraction\cite{zumbusch1940structures}}           \\
  Cu$_\text{4}$P$_\text{3}$       &        &          49          &        $P4/nmm$        &             AIRSS              \\
\rowcolor{gray!20}        CuP$_\text{2}$       &   $\star$   &          -           &       $P2_1/c$        &           ICSD 35282\footnote{Structure from X-ray diffraction \cite{mouller1982darstellung}}           \\
        CuP$_\text{3}$           &         &          26          &         $Pmmn$         &             AIRSS              \\
  Cu$_\text{8}$P$_\text{27}$    &          &          29          &          $Cm$          &               GA               \\
  Cu$_\text{8}$P$_\text{27}$    &          &          43          &          $Cm$          &               GA               \\
\rowcolor{gray!20}  Cu$_\text{2}$P$_\text{7}$   &  $\star$ &          -           &        $C2/m$         &           ICSD 35281\footnote{Structure from X-ray diffraction \cite{mouller1982darstellung}}           \\
\rowcolor{gray!50}  Cu$_\text{3}$P$_\text{11}$       &       &          17          &          $Cm$          &          Prototype 26563\footnote{Prototype structure $Cm$--Ag$_3$P$_{11}$ by single crystal X-ray diffraction \cite{MoellerInorg1981}}           \\
  Cu$_\text{7}$P$_\text{27}$    &          &          33          &          $Cm $         &               GA               \\
        CuP$_\text{4}$          &          &          32          &          $Cm$          &               GA               \\
\rowcolor{gray!20}       CuP$_\text{10}$     &     $\star$   &          -           &        $P\bar{1}$          &          ICSD 418805\footnote{Structure from single crystal X-ray diffraction \cite{CuP10}}           \\
\rowcolor{gray!20}              P           &      $\star$   &          -           &        $Cmca$         &          ICSD 150873\footnote{Black phosphorus structure from powder X-ray diffraction\cite{crichton2003phosphorus}}           \\
\end{tabular}

\end{minipage}

  \caption{Phases of Cu--P with formation energy $\leq$ 0\,meV/atom relative to Cu and P, and the distance from the convex hull tie-line, $\Delta E$, less than 50\,meV/atom.}
    \label{tab:CuP-0K0GPa}
\end{table}

 Details of the 24 structures which are both negative in formation energy relative to Cu and P, and are within 50\,meV/atom of the convex hull are given in Table \ref{tab:CuP-0K0GPa}. Phases on the convex hull tie-line in Figure \ref{fig:CuP-convex-hull} are indicated with $\star$ in Table \ref{tab:CuP-0K0GPa} and phases which are experimentally confirmed are highlighted in light gray. Phases not reported previously, within 20\,meV/atom of the convex hull tie-line, are highlighted in dark gray in Table \ref{tab:CuP-0K0GPa}. The provenance of each phase is given in the last column of Table \ref{tab:CuP-0K0GPa}. Phases from the ICSD are denoted with their ICSD Collection Code as ``ICSD \#'' . Phases which were found by swapping the elements of a prototype ICSD structure are denoted by the ICSD structure of the prototype used as ``Prototype \#''. 

 Of the 24 binary structures in Table \ref{tab:CuP-0K0GPa}, 9 were discovered by AIRSS, 6 by the GA, 4 from structural prototyping, and 4 were previously known Cu--P structures from the ICSD. Of particular interest are three new phases, highlighted in dark gray in Table \ref{tab:CuP-0K0GPa}, $Fm\bar{3}m$--Cu$_2$P, $I\bar{4}3d$--Cu$_3$P and $Cm$--Cu$_3$P$_{11}$ which are all within 20\,meV/atom of the convex hull and will be discussed further in the following sections.

\subsection{Phosphorus rich phase $Cm$--Cu$_3$P$_{11}$} \label{sec:highP}

\begin{figure}
\centering
\subfloat[$C2/m$--Cu$_2$P$_7$ unit cell]{\includegraphics[width=0.45\textwidth]{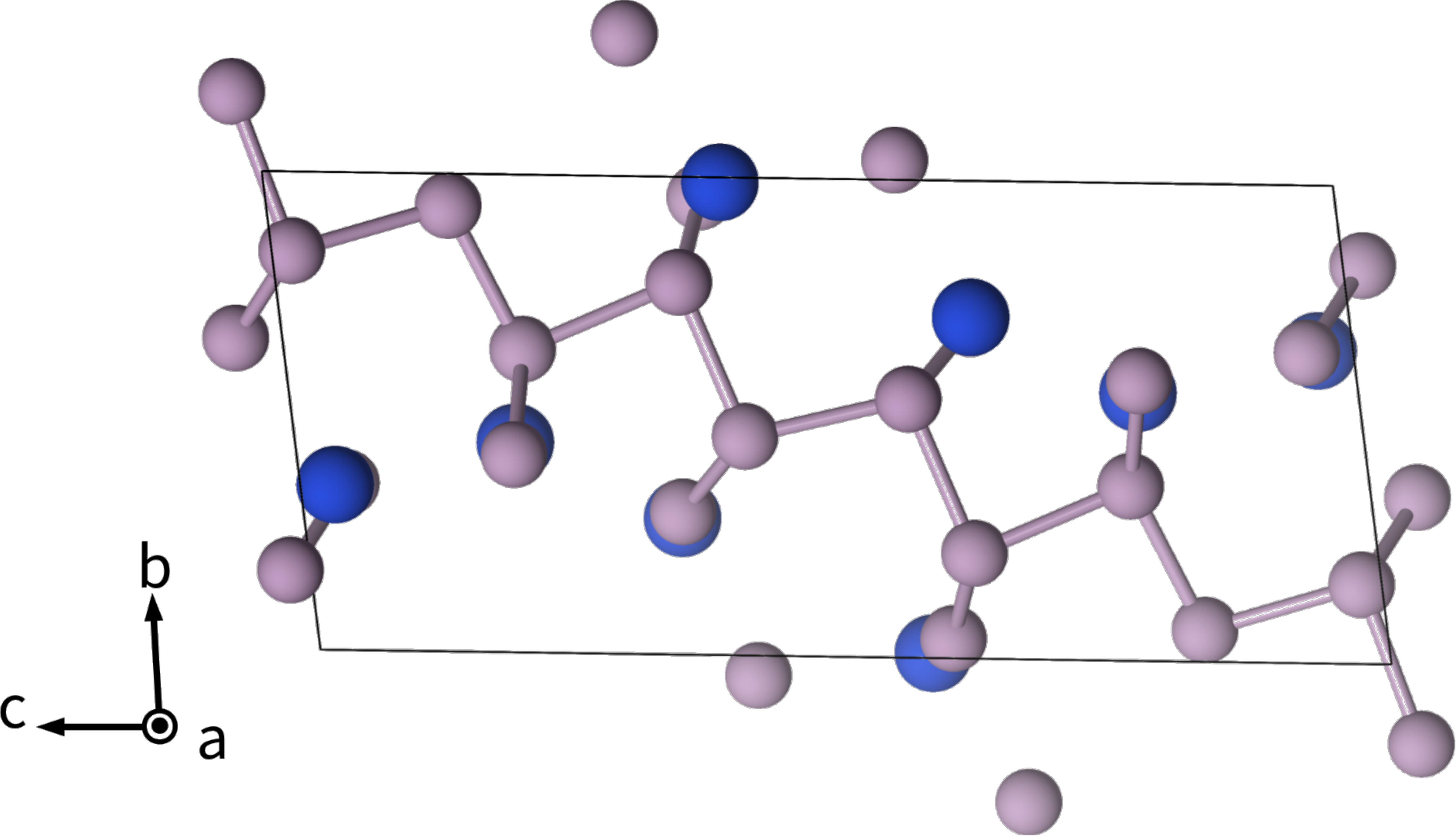}}
\qquad
\subfloat[$Cm$--Cu$_3$P$_{11}$ unit cell]{\includegraphics[width=0.45\textwidth]{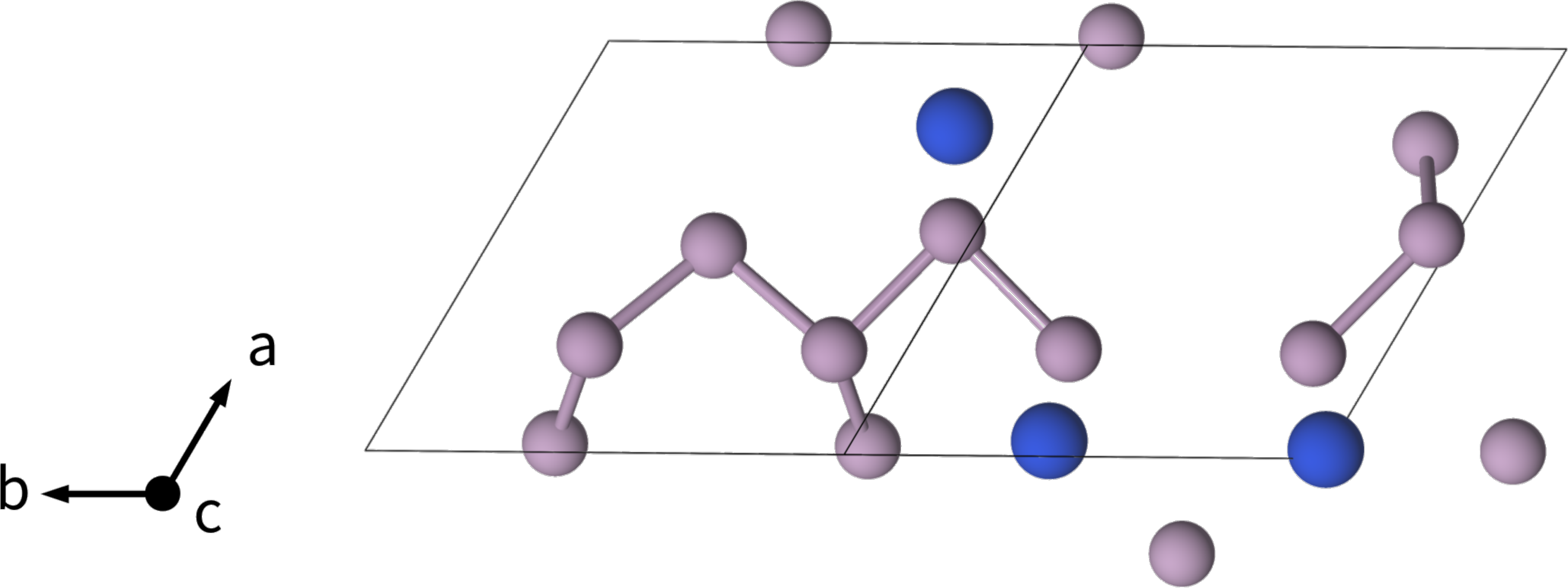}}
\qquad
\subfloat[$C2/m$--Cu$_2$P$_7$]{\includegraphics[width=0.45\textwidth]{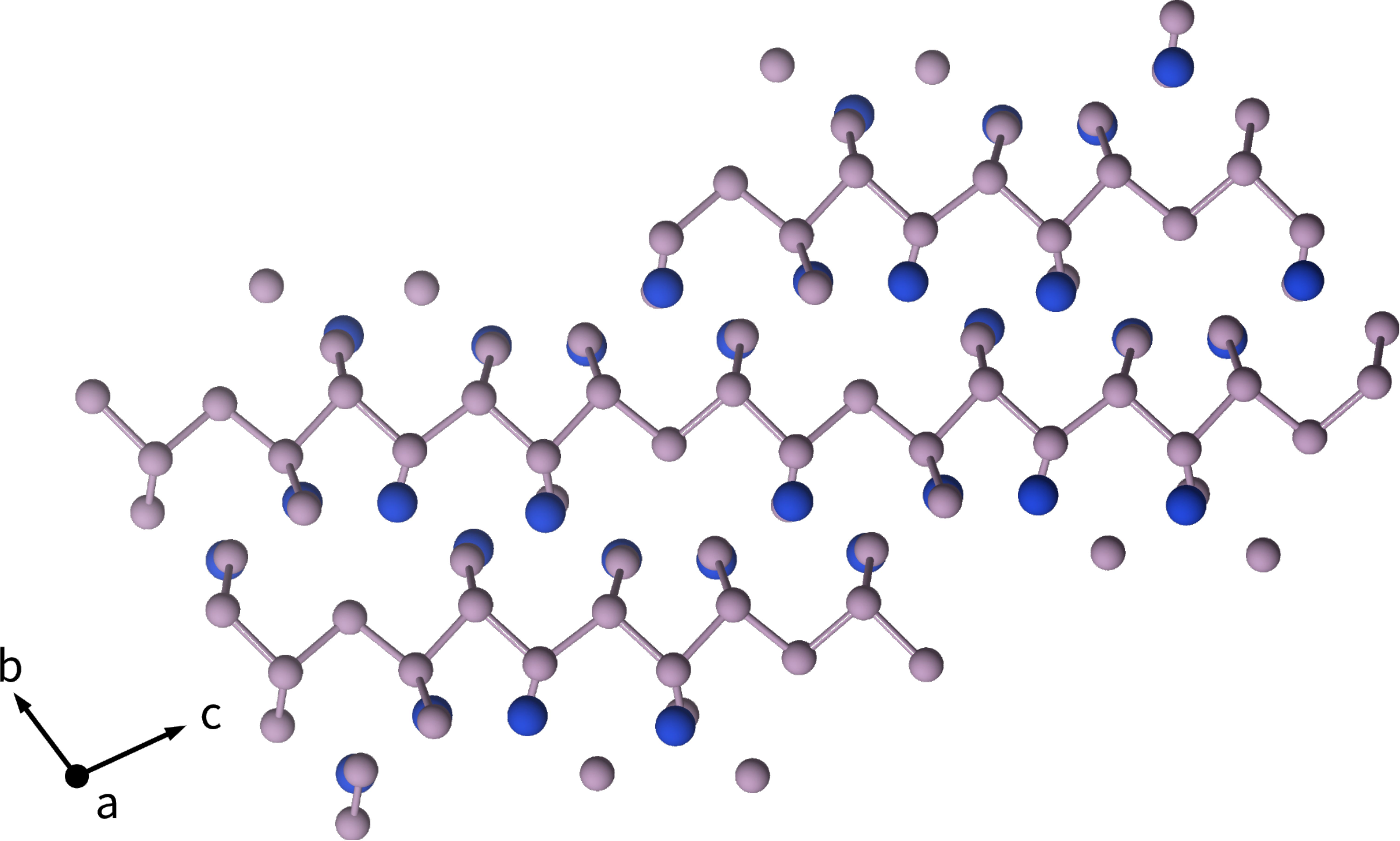}}
\qquad
\subfloat[$Cm$--Cu$_3$P$_{11}$]{\includegraphics[width=0.45\textwidth]{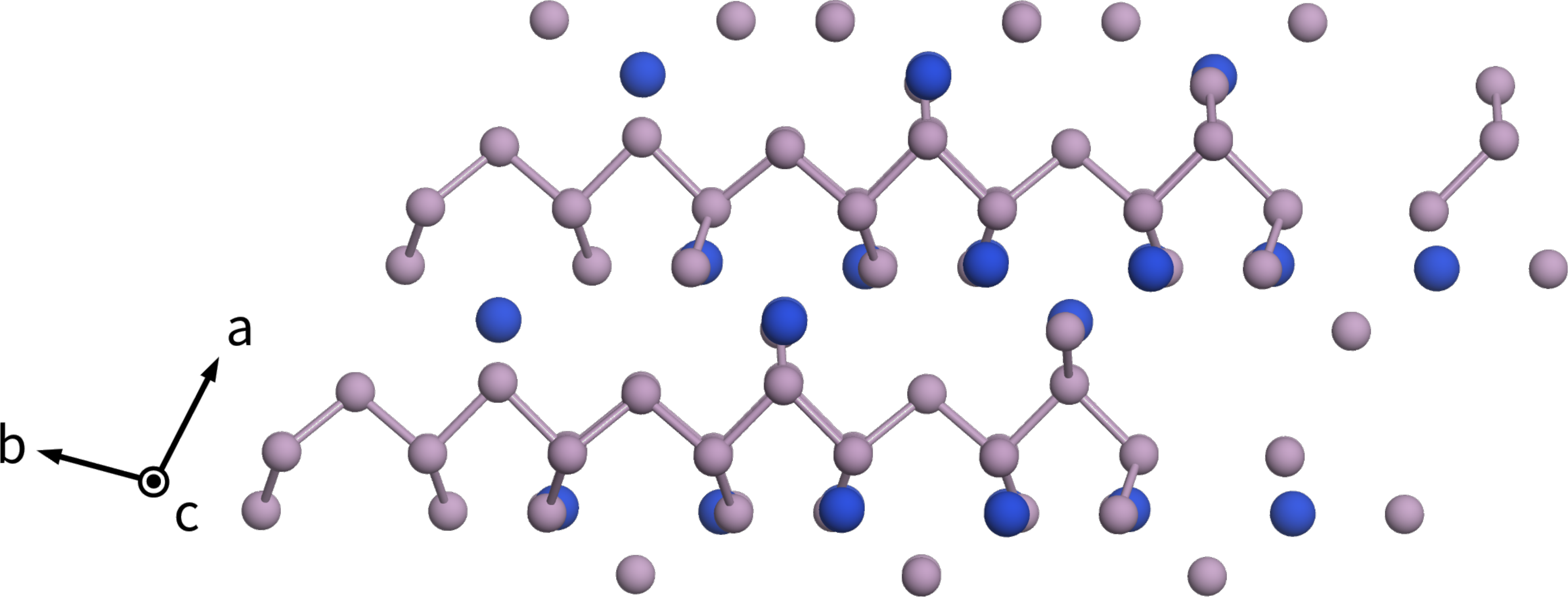}}

\caption[]{(a) Cu$_2$P$_7$ unit cell (b) Cu$_3$P$_{11}$ unit cell (c) Cu$_2$P$_7$ 2$\times$2$\times$2 supercell in which P--P connectivity is shown to highlight the P chains in the supercell structure. (d) Cu$_3$P$_{11}$ 2$\times$2$\times$2 supercell with P--P connectivity shown to show P chains as in the Cu$_2$P$_7$ supercell.}
\label{fig:Cu2P7Cu3P11struct}
\end{figure}

$Cm$--Cu$_3$P$_{11}$ is a new structure which was found by relaxing the prototype Ag$_3$P$_{11}$ (ICSD 26563); it is 17\,meV/atom from the hull tie-line, and has structural similarity to the ICSD structure $C2/m$--Cu$_2$P$_7$ (ICSD 35281)\cite{mouller1982darstellung} as shown in Figure \ref{fig:Cu2P7Cu3P11struct}. Both of these structures have repeating chains of P atoms, as seen in the supercells in Figure \ref{fig:Cu2P7Cu3P11struct}, in which alternating patterns of Cu or Cu--P are connected to a zig-zag chain of P atoms. All known phases in the P-rich (Cu$_x$P where $x<1$) region of the convex hull, namely $C2/m$--Cu$_2$P$_7$ (ICSD 35281) \cite{mouller1982darstellung}, $P2_1/c$--CuP$_2$ (ICSD 35281) \cite{mouller1982darstellung}, and $P\bar{1}$--CuP$_{10}$ (ICSD 418805), \cite{CuP10} have long chains of P atoms, similar to the layered $P_12/c_1$-P (ICSD 29273\cite{Thurna06597}, red P). 

\begin{figure}
  \centering
  \subfloat[]{\includegraphics[scale=1]{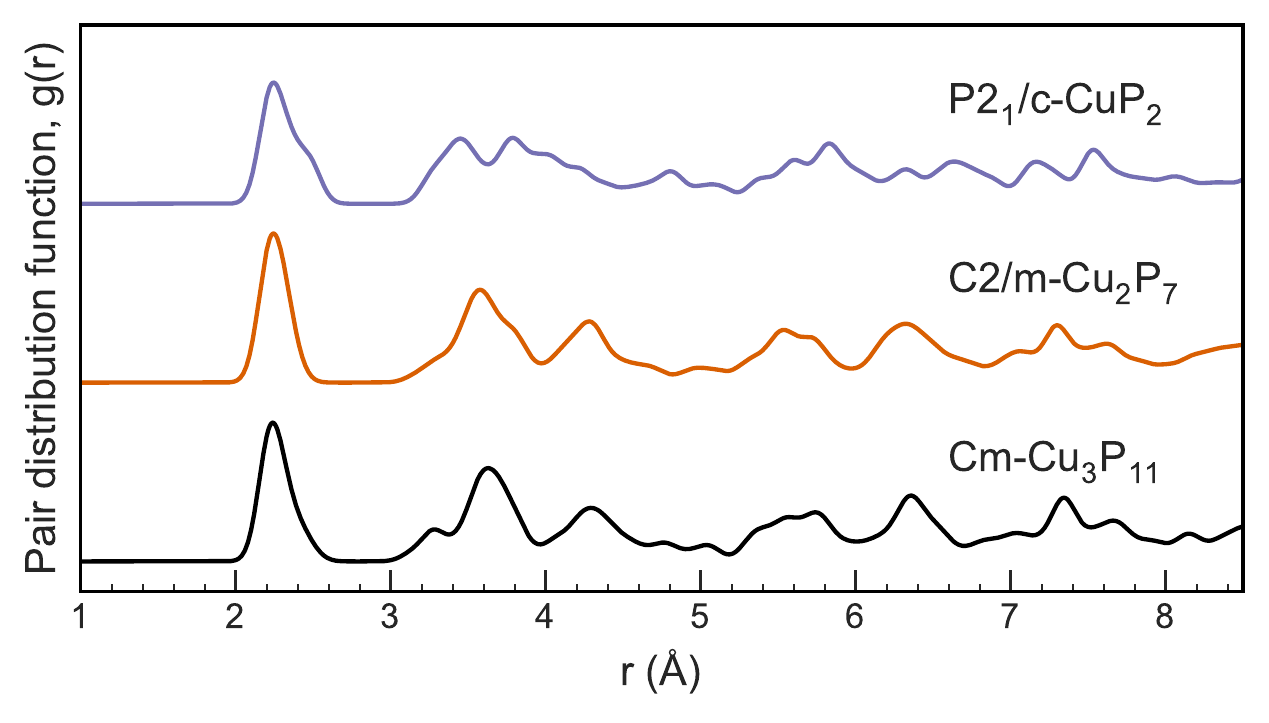}}
  
  \subfloat[]{\includegraphics[scale=1]{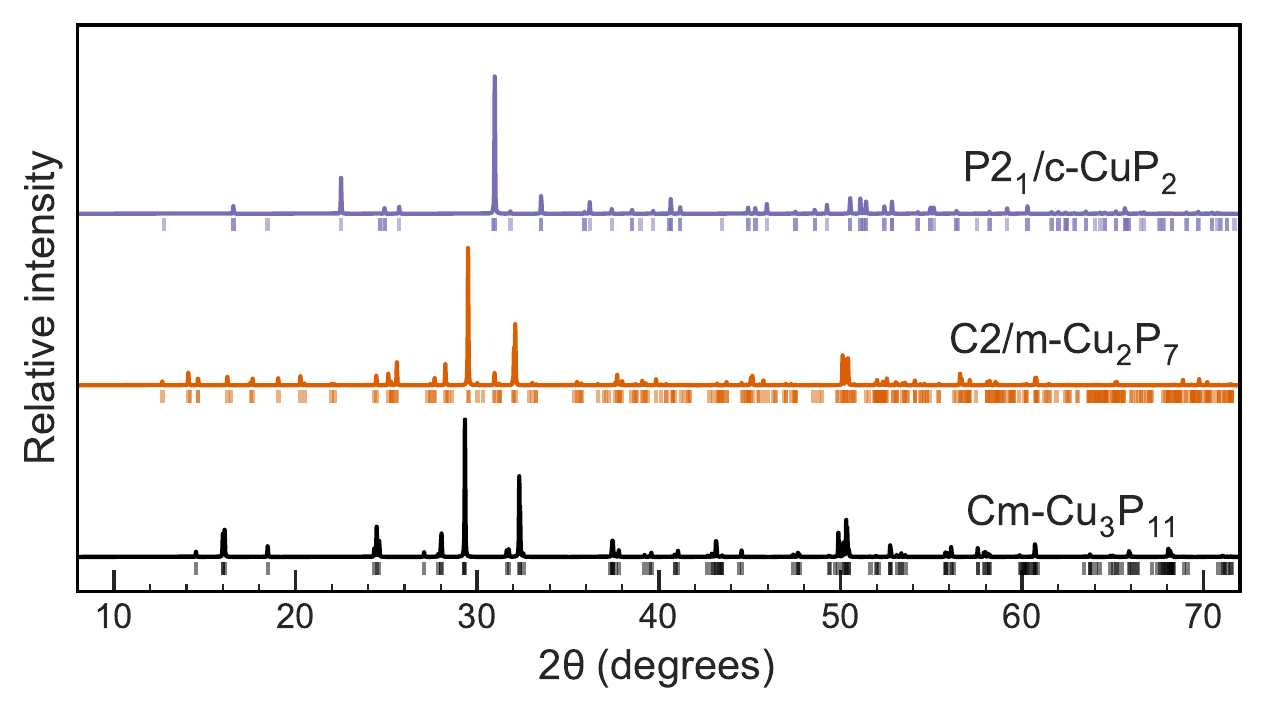}}
  \caption{(a) Pair distribution function (PDF) of $Cm$--Cu$_3$P$_{11}$, $P2_1/c$--CuP$_2$ (ICSD 35282), and $C2/m$--Cu$_2$P$_7$ (ICSD 35281) shows all three have a first peak between 2.20\,\AA\, and 2.24\,\AA, while CuP$_2$ has two peaks around 3.6\,\AA, where both Cu$_2$P$_7$ and Cu$_3$P$_{11}$ have one. All PDFs are artificially broadened with Gaussians of width 0.1\,\AA\,and PXRDs are calculated using a Cu K$_{\alpha}$ source. (b) The simulated PXRD patterns of both Cu$_2$P$_7$ and Cu$_3$P$_{11}$ share peak positions, as is expected from their shared symmetries. The Cu$_3$P$_{11}$ phase could be identified experimentally by the higher intensity peaks at 2$\theta < 30\degree$, including a distinct peak at 16$\degree$, not present in Cu$_2$P$_7$. Details of PXRD calculations can be found in the Supporting Information}.
  \label{fig:Cu3P11-pdf}
\end{figure}

 In the P-rich region, 5 new phases were identified within 50\,meV/atom of the convex hull: $Pmmn$--CuP$_3$, $Cm$--Cu$_8$P$_{27}$, $Cm$--Cu$_3$P$_{11}$, $Cm$--Cu$_{7}$P$_{27}$ and $Cm$--CuP$_4$. Using the GA it was possible to include structures with stoichiometries of P up to 27 atoms in the unit cell, and thus found structural variations on Cu$_2$P$_7$ such as Cu$_3$P$_{11}$. To compare the new metastable $Cm$--Cu$_3$P$_{11}$ structure with other P-rich structures, the pair distribution functions (PDF) and calculated powder x-ray diffraction (PXRD) peaks of CuP$_2$, Cu$_2$P$_7$ and Cu$_3$P$_{11}$ were calculated and are compared in Figure \ref{fig:Cu3P11-pdf}. In all three cases, the initial sharp peak in the PDF between 2.20 and 2.24\,\AA\,shows, unsurprisingly, the same Cu--P and P--P distance shared by all three structures. The peaks at radii above 3\,\AA\,show the longer range similarity between Cu$_3$P$_{11}$ and Cu$_2$P$_7$ which is not shared by CuP$_2$. Comparing the PXRD patterns of $C2/m$--Cu$_2$P$_7$ and $Cm$--Cu$_3$P$_{11}$ show that $Cm$--Cu$_3$P$_{11}$ is distinguished by a peak at a 2$\theta$ value of 16$\degree$, where $C2/m$--Cu$_2$P$_7$ has an indistinguishable peak at this point. Given the shared symmetry operations between $Cm$ and $C2/m$ we expect to see peaks at the same 2$\theta$ values, but the intensities will vary between the structures. We deduce that these three phases could be verified using experimental PXRD, by using the peaks at 2$\theta < 30\degree$ to distinguish between the phases. 

\subsection{Cu$_{3-x}$P phases ($x \leq 1$)}
\label{sec:Cu3xPPhases}

Within the stoichiometry range Cu$_{3-x}$P ($x \leq 1$), 4 unique Cu$_3$P phases, Cu$_{17}$P$_6$, Cu$_8$P$_3$, Cu$_7$P$_3$ and Cu$_2$P were found. Of these, $P6_3cm$--Cu$_3$P was the only phase previously experimentally determined, and had a formation energy 37\,meV/atom above the convex hull tie-line. Olofsson identified the stoichiometry of $P6_3cm$--Cu$_3$P at 975\,K to be between Cu$_{2.867}$P and Cu$_{2.755}$P due to Cu vacancies within the unit cell of $P6_3cm$--Cu$_{18}$P$_6$ (shown in Figure S1) \cite{olofsson1972crystal}. A study on low-temperature phases of Cu$_{3-x}$P proposes phases from Cu$_{2.3}$P to Cu$_{2.9}$P \cite{wolff2018low}. The lowest energy Cu$_{3-x}$P ($x \leq 1$) phases identified in Table \ref{tab:CuP-0K0GPa}, $P1$--Cu$_{17}$P$_6$ (Cu$_{2.83}P$), $Cmc2_1$--Cu$_8$P$_3$ (Cu$_{2.66}P$), and $P1$--Cu$_7$P$_3$ (Cu$_{2.33}P$) are all defect structures of $P6_3cm$--Cu$_3$P with 1, 2, and 4 Cu vacancies respectively from the $P6_3cm$--Cu$_{18}$P$_6$ unit cell of Cu$_3$P. Of these three $P6_3cm$--Cu$_3$P defect structures, $Cmc2_1$--Cu$_8$P$_3$ (Cu$_{2.66}P$) has the smallest distance from the hull ($\Delta E = 26$\,meV/atom). This corroborates previous DFT calculations suggesting Cu$_3$P has two Cu vacancies \cite{DeTrizio2015}.

\begin{figure}
  \centering
  \subfloat[$P2_1/m$--Cu$_3$P]{{\includegraphics[width=0.45\textwidth]{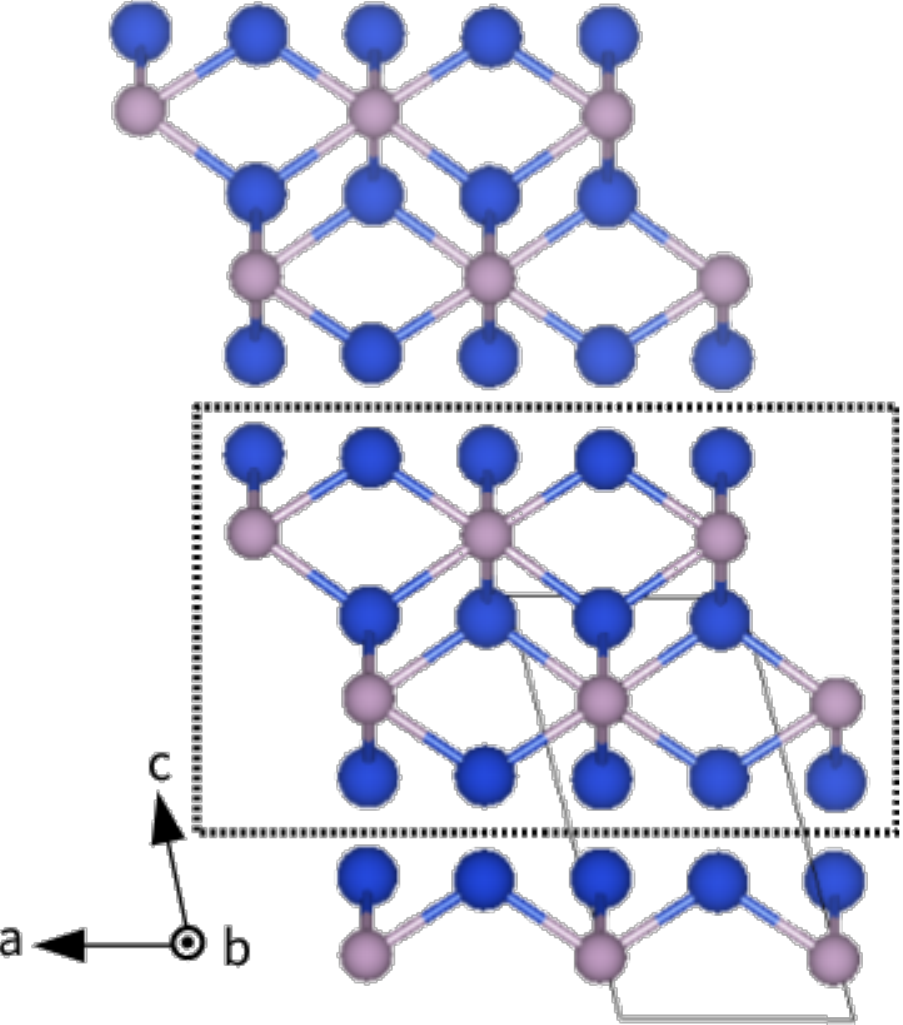} }}%
  \qquad
  \subfloat[$Fm\bar{3}m$--Cu$_2$P]{{\includegraphics[width=0.45\textwidth]{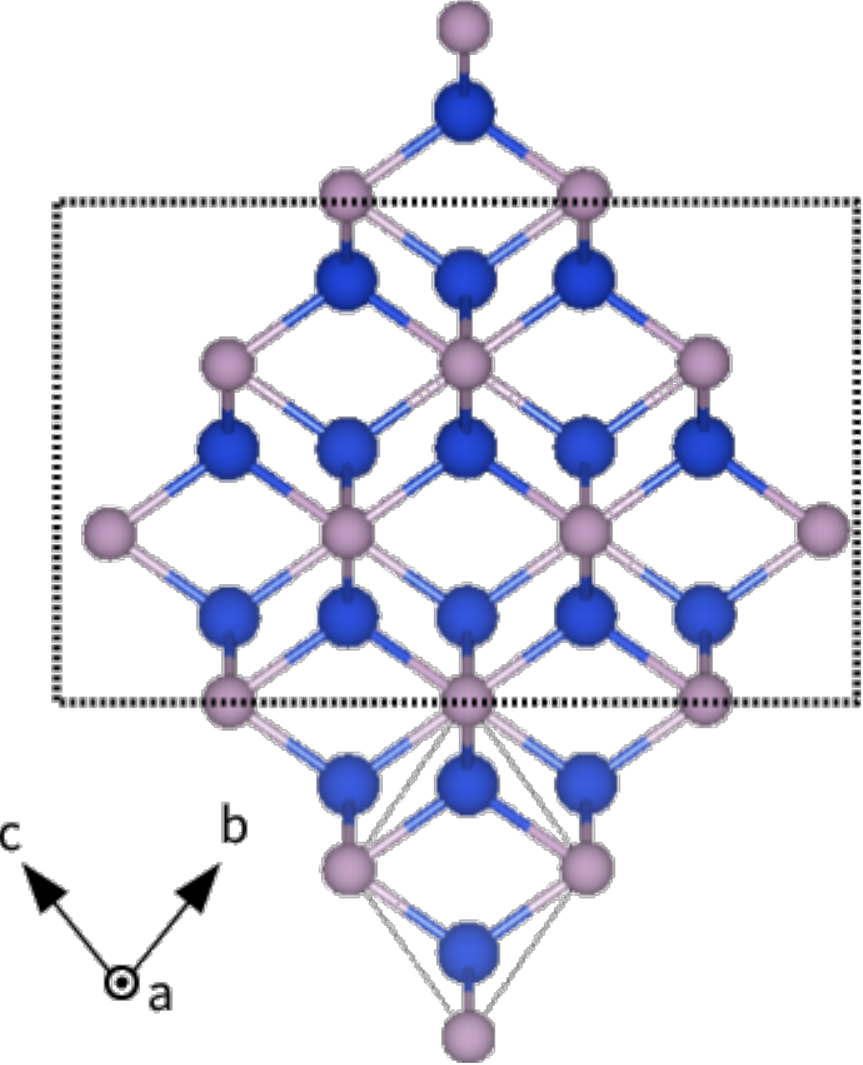} }}%
  \caption{$P2_1/m$--Cu$_3$P found by the GA and $Fm\bar{3}m$--Cu$_2$P from a swap with $Fm\bar{3}m$-Ir$_2$P. Here Cu atoms are colored blue and P atoms are colored pink. $P2_1/m$--Cu$_3$P structure can be described as a stacking of Cu$_2$P layers separated by Cu atoms. The stacking pattern which is present in both structures is indicated by the black dashed line surrounding the atoms in each structure. This is meant to guide the eye, and show the similarity in the two structures.  The unit cells of each structure are outlined with a thin gray box. Cu$_2$P is predicted to be a stable 2D phase \cite{C6CP01860B}, which could be layered to produce the Cu$_3$P phase shown in panel (a) here.}%
  \label{fig:Cu2PCu3Pstructs}%
\end{figure}

In addition to the ICSD phase of $P6_3cm$--Cu$_3$P ($\Delta E = 37$\,meV/atom), two other Cu$_3$P phases were found which are closer to the convex hull tie line than $P6_3cm$--Cu$_3$P; these are the $P2_1/m$--Cu$_3$P ($\Delta E = 30$\,meV/atom) and $I\bar{4}3d$--Cu$_3$P phase ($\Delta E = 11$\,meV/atom). The $P2_1/m$--Cu$_3$P phase is structurally related to the $Fm\bar{3}m$--Cu$_2$P ($\Delta E = 0$\,meV/atom) phase (discussed in the following section). These two phases are shown in Figure \ref{fig:Cu2PCu3Pstructs}, in which the $P2_1/m$--Cu$_3$P can be described as a stacking of the $Fm\bar{3}m$--Cu$_2$P phase. While the $Fm\bar{3}m$--Cu$_2$P phase has not been observed experimentally, it is likely that the two phases could be distinguished, given their distinct PDF and PXRD patterns shown in Figure S2. The PXRD pattern for $P2_1/m$--Cu$_3$P has additional low intensity peaks to the right of the 46$\degree$ peaks, and is distinct from the other low-energy phases of Cu$_3$P as shown in Figure S2, which would further distinguish this phase in experiment.

The lowest energy Cu$_3$P phase is an $I\bar{4}3d$ phase 11\,meV/atom above the tie-line, which was identified by relaxing the prototype $I\bar{4}3d$--Cu$_3$As structure (ICSD 64715 \cite{steenberg1938crystal,iglesias1977refinement}). The $I\bar{4}3d$--Cu$_3$P structure is the highest symmetry Cu$_3$P phase, and is the only cubic phase in the set of low-energy Cu$_3$P structures. $I\bar{4}3d$--Cu$_3$P contains 8 formula units in the primitive unit cell, and has 9-fold coordinated P atoms whereas $P6_3cm$--Cu$_3$P has 8-fold coordinated P atoms. The resulting crystal structures, shown in Figure S1, show two different long range orderings of the Cu sub-network. $P6_3cm$--Cu$_3$P has only one, 8-fold coordinated, P site which results in continuous zig-zag chains of Cu atoms surrounding the P, which are at the peaks of the buckles in the zig-zag. In $I\bar{4}$--Cu$_3$P, there are two 9-fold coordinated sites; one site at the center of the surrounding Cu (seen in Figure S1) and one at the edges, which together form a hexagonal Cu cage surrounding the P atom in the center. While both phases have high-coordinated P atoms, the $I\bar{4}3d$--Cu$_3$P shows a network of Cu atoms surrounding a central P atom, where $P6_3cm$--Cu$_3$P contains infinite Cu chains in the $c$ direction.

Another trigonal phase, $P\bar{3}c1$--Cu$_3$P (ICSD 16841\cite{steenberg1938crystal}, $\Delta E > 50$\,meV/atom) has the same structure as $P\bar{3}c1$--Cu$_3$As (ICSD 16840 \cite{steenberg1938crystal}), however it is 82\,meV/atom above the convex hull tie-line. To the best of our knowledge, there are no reports of an $I\bar{4}3d$--Cu$_3$P phase, either experimentally or in a computational database. The PDF and PXRD patterns of $I\bar{4}3d$--Cu$_3$P given in Figure S2, show no relation to any other Cu$_3$P phase, or the $Fm\bar{3}m$--Cu$_2$P phase, thus, if energetically stable, it could be identified using PXRD in experiment.

\subsection{$Fm\bar{3}m$--Cu$_2$P}
\label{sec:resultsCu2P}

The $Fm\bar{3}m$--Cu$_2$P phase was found from the prototype $Fm\bar{3}m$-Ir$_2$P (ICSD 640898) \cite{rundqvist1960phosphides}. Comparing the Cu$_2$P phase to both Ir$_2$P and Rh$_2$P using PDFs in Figure S3 shows that the PDFs are identical between all three structures, and the PXRD plot of Cu$_2$P has the same peaks, all shifted to slightly higher values of 2$\theta$ due to structural relaxations in the geometry optimization of Cu$_2$P.

Previously, a 2D structure of Cu$_2$P was predicted theoretically as a buckled non-magnetic material \cite{C6CP01860B}, in which the magnetism expected was inhibited by the buckled layers. The buckled layers from the 2D phase are also present in the bulk $Fm\bar{3}m$--Cu$_2$P, and the non-magnetic nature was confirmed in the bulk phase by the lack of spin-polarization in the density of states shown in Figure S4. The bulk $Fm\bar{3}m$--Cu$_2$P structure described above has the same structural motifs as the 2D hexagonal phase found by Yang et al.\cite{C6CP01860B}, and has the same electronic properties.

\begin{figure}
    \subfloat[]{{\includegraphics[width=0.9\textwidth]{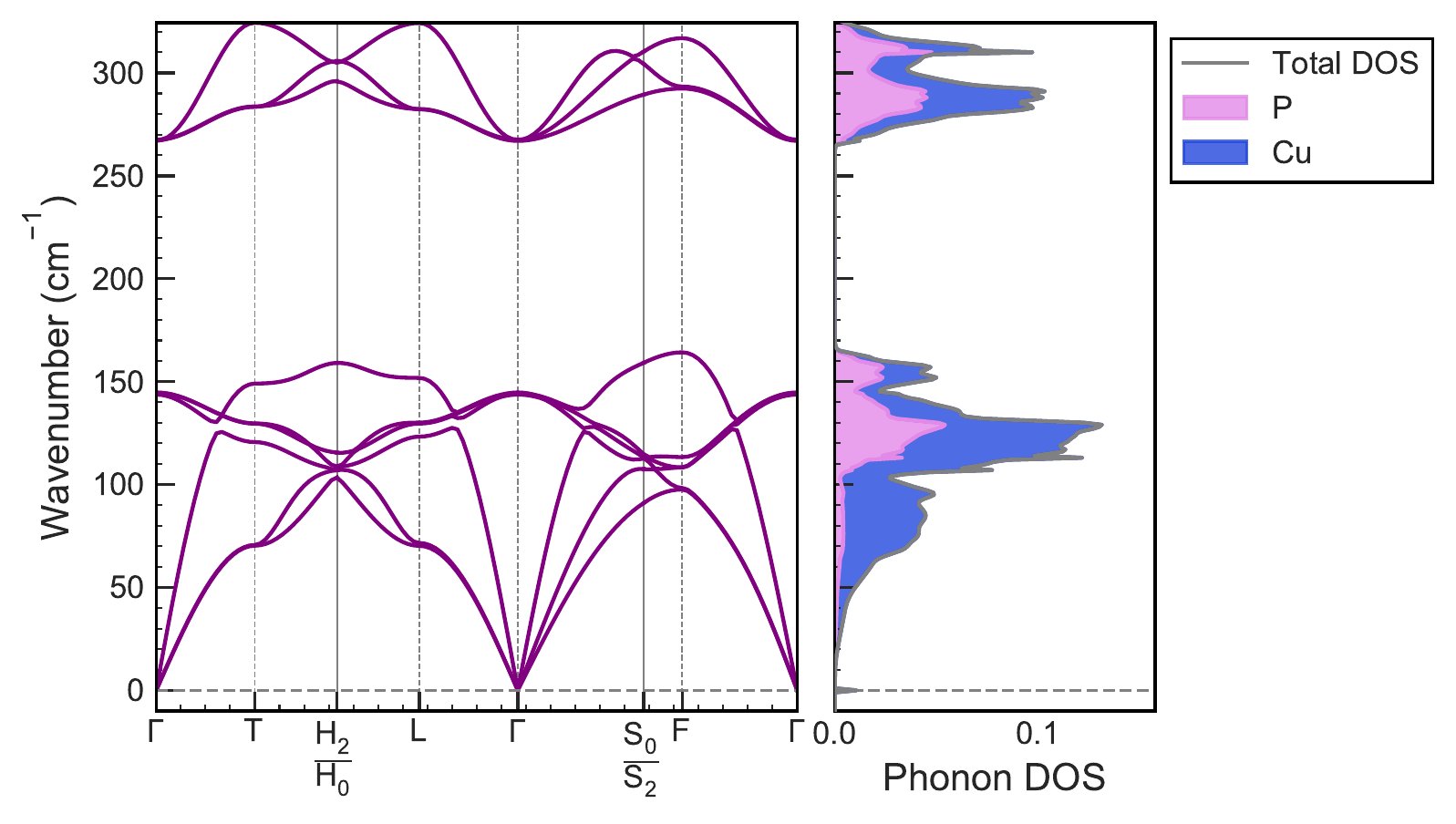}}}%
    
    \subfloat[]{{\includegraphics[width=0.8\textwidth]{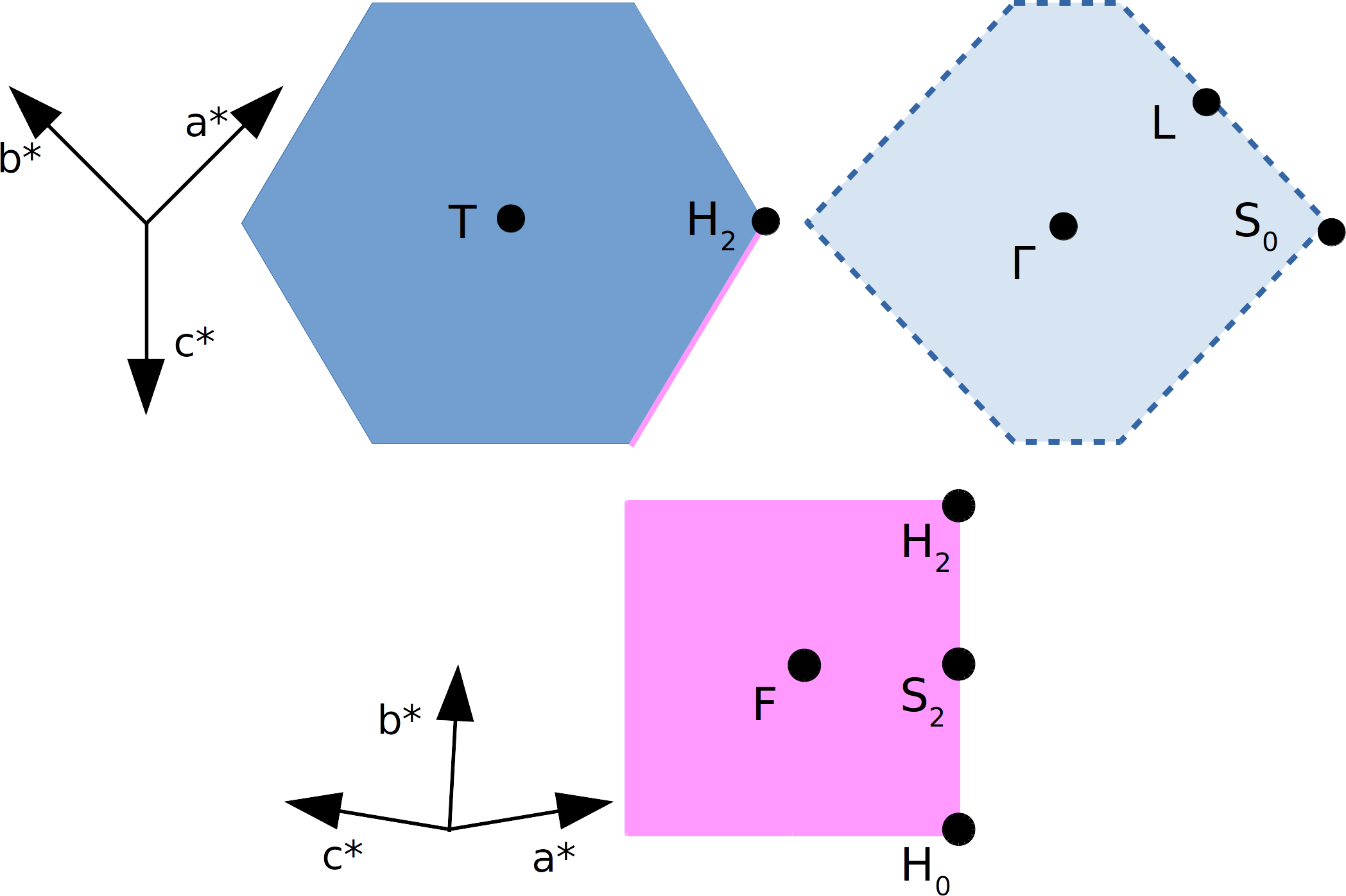} }}%
    \caption{(a) Cu$_2$P phonon dispersion under the harmonic approximation and the corresponding Brillouin zone. Phonon dispersion and density of states were interpolated from the dynamical matrix calculated using the PBE $xc$-functional and the ``C18'' pseudopotential library, with a 2$\pi\times$0.03\,\AA$^{-1}$ $k$-point spacing, 500\,eV plane wave cutoff, in a $2\times2\times2$ supercell using the finite displacement method (corresponding to a phonon $q$-point spacing of 2$\pi$\,$\times$\,0.046\,\AA$^{-1}$). (b) Brillouin zone is a truncated octahedron, with special points on each face labeled here. The dashed hexagon outline indicates the center of the Brillouin zone.}%
    \label{fig:Cu2P-phonons}%
  \end{figure}

$Fm\bar{3}m$--Cu$_2$P lies on the convex hull tie-line, and is energetically more stable than both the experimentally confirmed phase of $P6_3cm$--Cu$_3$P, and its defect structure $Cmc2_1$--Cu$_8$P$_3$. Figure \ref{fig:Cu2P-phonons} shows the phonon dispersion for the $Fm\bar{3}m$--Cu$_2$P computed as mentioned in the Methods section. No imaginary phonon frequencies were present in the dynamical matrix (interpolated or otherwise), indicating that $Fm\bar{3}m$--Cu$_2$P is dynamically stable.

\begin{figure}
  \centering
  \includegraphics[width=0.9\textwidth]{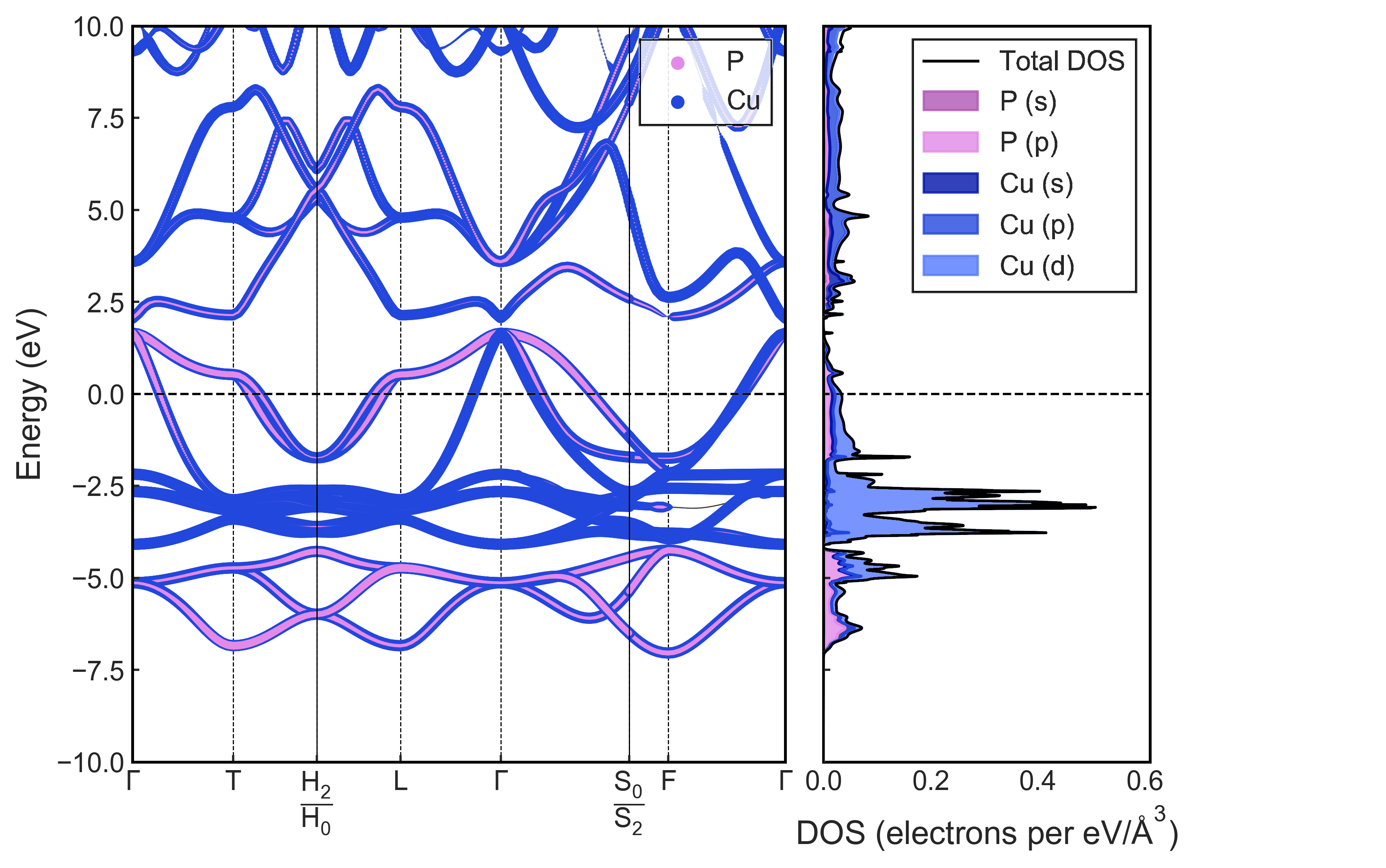}
  \caption{Electronic band structure of Cu$_2$P projected onto the Cu and P states and density of states projected onto the Cu $s$,$p$,$d$ and P $s$ and $p$ states, for Cu$_2$P using an energy cutoff of 500\,eV with 2$\pi\times0.03\,$\AA$^{-1}$ $k$-point spacing and ``C18'' on-the-fly pseudopotentials. The projected band structure is produced by OptaDOS \cite{morris2014optados}, and band energies are calculated by CASTEP \cite{clark2005first}.}
  \label{fig:Cu2P-bands}
\end{figure}

The electronic structure of $Fm\bar{3}m$--Cu$_2$P is related to the electronic structure of other $Fm\bar{3}m$ TMPs, suggesting it belongs to the same class of materials as $Fm\bar{3}m$-Ir$_2$P and Rh$_2$P. Of the TMPs in the Materials Project database \cite{Jain2013}, 21 are insulating, and 68 are metallic with a high density of transition metal $d$-bands below the Fermi level. Figure \ref{fig:Cu2P-bands} shows the electronic band structure and density of states of $Fm\bar{3}m$--Cu$_2$P projected by species along the high-symmetry path from SeeK-path used previously, and the density of states projected by angular momentum channel on a fine Monkhorst-Pack grid. The band structure shows that Cu$_2$P is a metal with P and Cu bands touching at the $\Gamma$ point \textasciitilde2.0\,eV above the Fermi level. In addition there is a characteristic high density of flat bands localized on the Cu ions that exhibit $d$-character around 2.5\,eV below the Fermi level. Calculating this band structure using the HSE06 functional (shown in Figure S5), a hybrid functional designed to correct for band gap underestimation, the gap at $\Gamma$ between the Cu and P bands is closed. 

Many M$_2$P phases (where M is a transition metal) have a structure similar to $P\bar{6}2m$-Ni$_2$P\cite{larsson1965x} and Fe$_2$P \cite{carlsson1973determination,song2009nature} in which the metal atoms sit in a cage of 3-fold coordinated P and 4-fold coordinated metal atoms. $Fm\bar{3}m$--Cu$_2$P is most similar to the other $Fm\bar{3}m$ TMPs, as it was derived from a the prototype structure $Fm\bar{3}m$-Ir$_2$P, and has 4-fold coordinated Cu with 8-fold coordinated P. The Cu$_2$P band structure in Figure \ref{fig:Cu2P-bands} is also similar to those of Ir$_2$P and Rh$_2$P. In Rh$_2$P there is a directionally opened gap 1\,eV above the Fermi level at the $\Gamma$ point, not present in Cu$_2$P or Ir$_2$P (Cu$_2$P, Ir$_2$P and Rh$_2$P band structures calculated with spin-orbit coupling are given in Figure S6, the gapped region is outlined in black dashed line). Both of these structures exhibit spin-orbit coupling due to their heavy metal ions, while Cu has negligible spin-orbit coupling effects. The Rh$_2$P and Ir$_2$P band structures are calculated including spin-orbit coupling.

\subsection{Finite-Temperature Phase Stability}
\label{resultsTempHull}

The temperature-dependent convex hull was constructed by calculating the finite-temperature Gibbs free energies by including vibrational effects at the harmonic level \cite{Baroni2001} of several related structures on or near the convex hull from Figure \ref{fig:CuP-convex-hull}. All structures within 20\,meV/atom of the hull at 0\,K were included in the finite-temperature hull; these were $Fm\bar{3}m$--Cu$_2$P, $I\bar{4}3d$--Cu$_3$P, $Cmc2_1$--Cu$_8$P$_3$ (the structure with 2 Cu vacancies from $P6_3cm$--Cu$_3$P discussed previously), CuP$_2$, CuP$_{10}$, Cu$_2$P$_7$, and Cu$_3$P$_{11}$.
  
  \begin{figure}
      \centering
      \subfloat[]{{\includegraphics[scale=0.9]{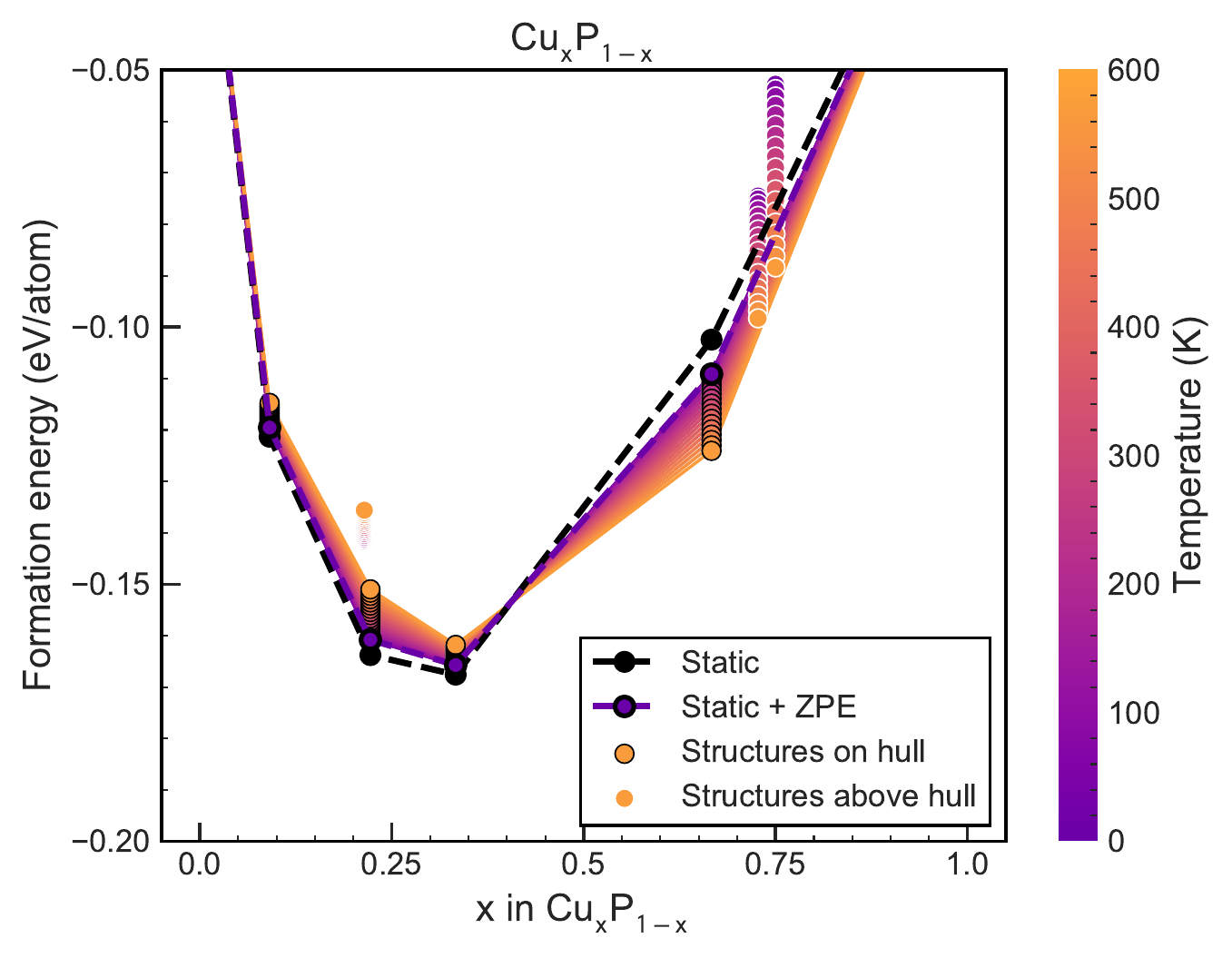}}}%
      
      \subfloat[]{{\includegraphics[scale=0.9]{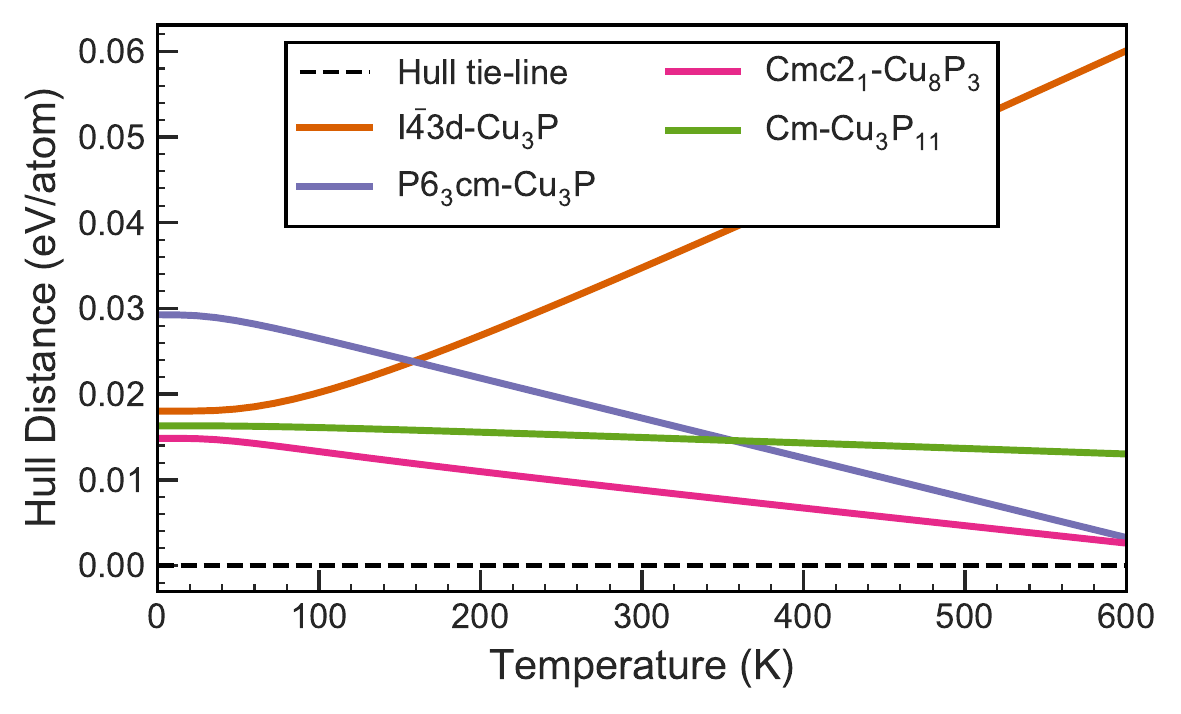} }}%
      \caption{(a) Temperature-dependent convex hull with PBE $xc$-functional and MBD correction on the black phosphorus chemical potential. (b) Distance from the hull for the four structures which were above the convex hull tie-line at 0\,K. All structures except $I\bar{4}3d$--Cu$_3$P get closer to the tie line as the temperature increases, suggesting they are stabilized by temperature. This agrees with the experimental evidence for Cu$_8$P$_3$ and $P6_3cm$--Cu$_3$P, and suggests that Cu$_3$P$_{11}$ may form experimentally. }%
      \label{fig:CuP-thermo-hull}%
  \end{figure}

  The chemical potentials for this binary convex hull were $Cmca$-P (black phosphorus) and $Fm\bar{3}m$--Cu. Previously, Mayo \textit{et al.} noted that the inclusion of semi-empirical dispersion corrections for black phosphorus changed the energetics of the convex hull \cite{mayo2016ab}, and therefore it is not possible to combine optimized structures with and without dispersion corrections on the same convex hull. However, in order to obtain non-imaginary phonon frequencies of $Cmca$-P, it is necessary to account for dispersion. To account for this the many-body dispersion correction (MBD) was applied during the geometry optimization and phonon calculation \cite{tkatchenko2012accurate}. Using PBE the distance between P chains in black phosphorus is 3.95 \AA. By applying this correction, the P--P chain distance was reduced to 3.58 \AA. In order to include the MBD black phosphorus on the convex hull in Figure \ref{fig:CuP-thermo-hull}a, we calculated the free energy of black phosphorus $F(T)$, at a given temperature $T$ as,

   \begin{equation}
     F(T) = H + \Delta F^{MBD*}(T),%
   \end{equation}
  where $H$ is the enthalpy without the dispersion correction, and $\Delta F^{MBD*}(T)$ is the free energy contribution at temperature $T$ with the MBD dispersion correction, which includes the zero-point energy. In this way the energies of black phosphorus were referenced to the ground-state energy without dispersion. The SCAN functional accurately describes the phonon modes of black phosphorus without any added dispersion corrections (i.e. no imaginary modes are observed) and therefore is used as a comparison to the MBD corrected PBE functional in Figure S7. Figure S7 shows that for any temperature $T$, both $\Delta F^{MBD}(T)$ and $\Delta F^{SCAN}(T)$ are on the same scale, only the zero-point energy is shifted (by 2.6\,meV/atom) for the PBE+MBD calculation. Therefore, we expect the results of the PBE+MBD free energies to be comparable with non-dispersion corrected PBE free energies.
  
  Using the MBD correction on black phosphorus in addition to the phonon modes of the previously mentioned phases of Cu--P, the hull in Figure \ref{fig:CuP-thermo-hull}a was constructed up to 600\,K, above which no changes to stabilities are observed. A maximum value of 600\,K was chosen so as not to approach the melting point of any phases, as the known phases of Cu--P typically have melt between 800 and 1200\,K. Furthermore, the harmonic approximation is a limited approach, and at higher temperatures, anharmonicity should be accounted for. $Fm\bar{3}m$--Cu$_2$P remains on the hull at 600\,K, suggesting it could be synthesized at high temperature. The convex hull is confirmation that the $Cmc2_1$--Cu$_8$P$_3$ phase formed from two Cu vacancies in $P6_3cm$--Cu$_3$P is the more stable phase at room temperature, as at 300\,K, $Cmc2_1$--Cu$_8$P$_3$ is within 10\,meV/atom of the convex hull as shown in Figure \ref{fig:CuP-thermo-hull}b. In addition, the destabilization of $I\bar{4}3d$--Cu$_3$P at high temperatures, shown in Figure \ref{fig:CuP-thermo-hull}b suggests that this phase is not experimentally realizable, and provides an explanation as to why it has not yet been experimentally synthesized. We can clearly see that $P6_3cm$--Cu$_3$P is stabilized at higher temperatures, as shown in Figure \ref{fig:CuP-thermo-hull}b, in which it is within 10\,meV/atom of the convex hull at temperatures higher than 450\,K.
  
  Previous work on the Cu$_{3-x}$P phases of $P6_3cm$--Cu$_3$P \cite{DeTrizio2015} confirms that the formation of two vacancies in Cu$_3$P is energetically stabilizing.  By enumerating all of the possible structures with two Cu vacancies using the vacancy enumeration procedure described in the Methods Section, we have determined that the $Cmc2_1$--Cu$_8$P$_3$ phase with two Cu vacancies in the 6$c$ Wyckoff positions is the lowest energy vacancy phase.  Given the large number of ways to introduce these vacancies into the structure, configurational entropy will further stabilize this phase at high temperatures.  To fully understand the nature of vacancy formation in $P6_3cm$--Cu$_3$P a full cluster expansion could be performed, which is beyond the scope of this paper.
  
\subsection{Cu$_2$P as a Li-ion battery conversion anode}
\label{sec:battery}

$Fm\bar{3}m$--Cu$_2$P was computationally predicted to be energetically stable as both a 2D material \cite{C6CP01860B} and now in this article as a bulk phase. The previous sections predict the stability of $Fm\bar{3}m$--Cu$_2$P at temperatures up to 600\,K, and characterize it as a metal with dispersive bands and delocalized conduction states at the Fermi level. An intuitive choice of application for Cu$_2$P lies in conversion anodes for Li-ion batteries, where previously both CuP$_2$ and Cu$_3$P were used as anodes with gravimetric capacities between 300 and 800\,mAh/g \cite{Kim2017CuP2,WANG2003480,PFEIFFER2004263,BICHAT200480}.

The crystal structure of $P6_3cm$--Cu$_3$P has a theoretical capacity of 363\,mAh/g and experimentally has exhibited a range of capacities based upon the preparation method used \cite{BICHAT200480}. The powdered Cu$_3$P anodes prepared by Bichat et al\cite{BICHAT200480} ranged in initial capacity from 272\,mAh/g using high-temperature synthesis in a silica tube to 527\,mAh/g using low-temperature solvothermal synthesis respectively. In the solvothermal route, the Cu$_3$P powders were prepared with copper chloride, water, and NH$_4$OH with white phosphorus, which could have resulted copper oxide impurities leading to the initial capacity which is above the theoretical capacity of crystalline Cu$_3$P. Cu$_3$P powder synthesized by a solid-state reaction with red P in an ethanol suspension and Cu foil, the initial capacity of Cu$_3$P was 415\,mAh/g \cite{PFEIFFER2004263}. Energy dispersive X-ray analysis showed that the stoichiometry was close to Cu$_3$P (though not exact) suggesting that this initial structure could have been within the stoichiometric range of Cu$_{3-x}$P in order to achieve that initial capacity, in addition to the added capacity from likely oxide impurities. 

In contrast, Cu$_2$P has a theoretical capacity of 509\,mAh/g, which is above that of graphite at 372\,mAh/g. The metallic nature of Cu$_2$P further enhances its use as a Li-ion battery anode, enabling fast electronic transfer through the electrode of the battery. In fact, Cu is already widely used as a current collector in contemporary Li-ion batteries, and previous studies on Cu$_3$P nanorods suggest that Cu--P anodes create a synergistic chemical interface with the Cu-current collector which promotes cyclability \cite{villevieille2008good}. Furthermore, because of its comparatively lower P content, volume changes during cycling are reduced, and therefore the degradation is likely to be less severe \cite{WANG2003480}.

The volume expansion for a conversion anode, with an overall conversion reaction

\begin{equation}
  \text{Cu}_a\text{P}_b + 3b\,\text{Li} \longrightarrow a\,\text{Cu} + b\,\text{Li}_3\text{P},
  \label{eq:convrxn}
\end{equation}%
is calculated as,

\begin{equation}
  \mathrm{Volume\,expansion\,(\%)} = 100 \times \left(\frac{b\,V(\mathrm{Li_3P}) + a\,V(\mathrm{Cu})}{V(\mathrm{Cu}_a\mathrm{P}_b)} - 1\right),
  \label{eq:volexp}
\end{equation}%
where $V(\mathrm{Cu}_a\mathrm{P}_b)$ is the volume per formula unit of each phase in the conversion reaction. Using this equation, the volume expansion of $Fm\bar{3}m$--Cu$_2$P is 99\,\%. This is comparable to the calculated volume expansion of $P63_cm$--Cu$_3$P, which is 86\,\%, and far superior to the volume expansion of CuP$_2$, which is 165\,\%. The volume expansion for each binary Cu--P phase is shown in Figure S8, confirming that Cu$_2$P has the lowest volume expansion of the four stable phases on the convex hull. Experimental reports on cycling of ball-milled CuP$_2$ \cite{WANG2003480} suggest that volume expansion occurs, as after cycling for 10 cycles, the capacity is reduced by 50\,\%, although they give no estimate of the level of volume expansion in the cell. The expansion is partially mitigated through the use of nanostructuring \cite{Kim2017CuP2}, which allows cycling for 200 cycles. However, there is still capacity fading in this case, which reiterates the need for a high capacity conversion anode with low volume expansion, so as to reduce the need for nanostructuring or other post-processing techniques to mitigate volume expansion. As both Cu$_3$P and Cu$_2$P have lower predicted volume expansions, and synthesized Cu$_3$P shows no evidence of deleterious volume expansion \cite{BICHAT200480}, it is likely that Cu$_2$P would also have minimal volume expansion in experiment.

Using the convex hull constructed in Figure \ref{fig:CuP-convex-hull} and the structures on the ternary hull of Cu--Li--P, a voltage profile was constructed from the DFT ground-state energies for both $Fm\bar{3}m$--Cu$_2$P and $P6_3cm$--Cu$_3$P. All of the known ternary structures were included in this hull: $P\bar{3}m1$--Cu$_2$LiP, $I4/mmm$--Cu$_2$LiP$_2$, $Immm$--Cu$_4$Li$_5$P$_6$ and $Cmcm$--CuLi$_2$P, as well as the binary Li--P structures $Cmcm$-Li$_3$P, $P2_1/c$-LiP, $P2_12_12_1$-Li$_3$P$_7$ and $I4_1/acd$-LiP$_7$. A plane wave kinetic energy cutoff of 700\,eV was used, and all structures on the hull were re-relaxed at this higher cutoff. The ternary hull is shown in Figure S9, in which the pathways from Cu--P to Li are also shown, to depict how the voltage profiles for these Cu--P phases were calculated. The hull is shaded with a colormap to show the relative formation energy of phases on the hull, indicating that the Li--P phases have larger formation energies (and thus create a deeper convex hull) than the Cu--P phases. 

Although Cu--Li phases are predicted to be stable under the approximation of PBE, the formation energy of the predicted Cu$_3$Li phase is only 26\,meV/atom in the OQMD database \cite{saal2013materials,van2004first} and no phases of Cu--Li are predicted at finite temperature in experiment \cite{okamoto2011cu}. Furthermore, Cu is used as a current collector in Li ion batteries specifically for its properties in resisting Li intercalation, and dead Li is found during cycling rather than Cu--Li phases \cite{fang2019quantifying}. Therefore, no Cu--Li compounds were included in the convex hull. 

There are three ternary compounds on the Cu--Li--P hull in Figure S9; these are $I4/mmm$--Cu$_2$LiP$_2$, $Immm$--Cu$_4$Li$_5$P$_6$ and $Cmcm$--CuLi$_2$P. Experiments suggest that a hexagonal LiCu$_2$P phase forms \cite{Bichat_2004} during cycling, however the $P\bar{3}m1$--Cu$_2$LiP (ICSD 659706) \cite{SchlengerCu3P} phase of this structure is 39\,meV/atom above the hull at a plane wave cutoff of 700\,eV.

\begin{figure}
  \centering
  \includegraphics[scale=1]{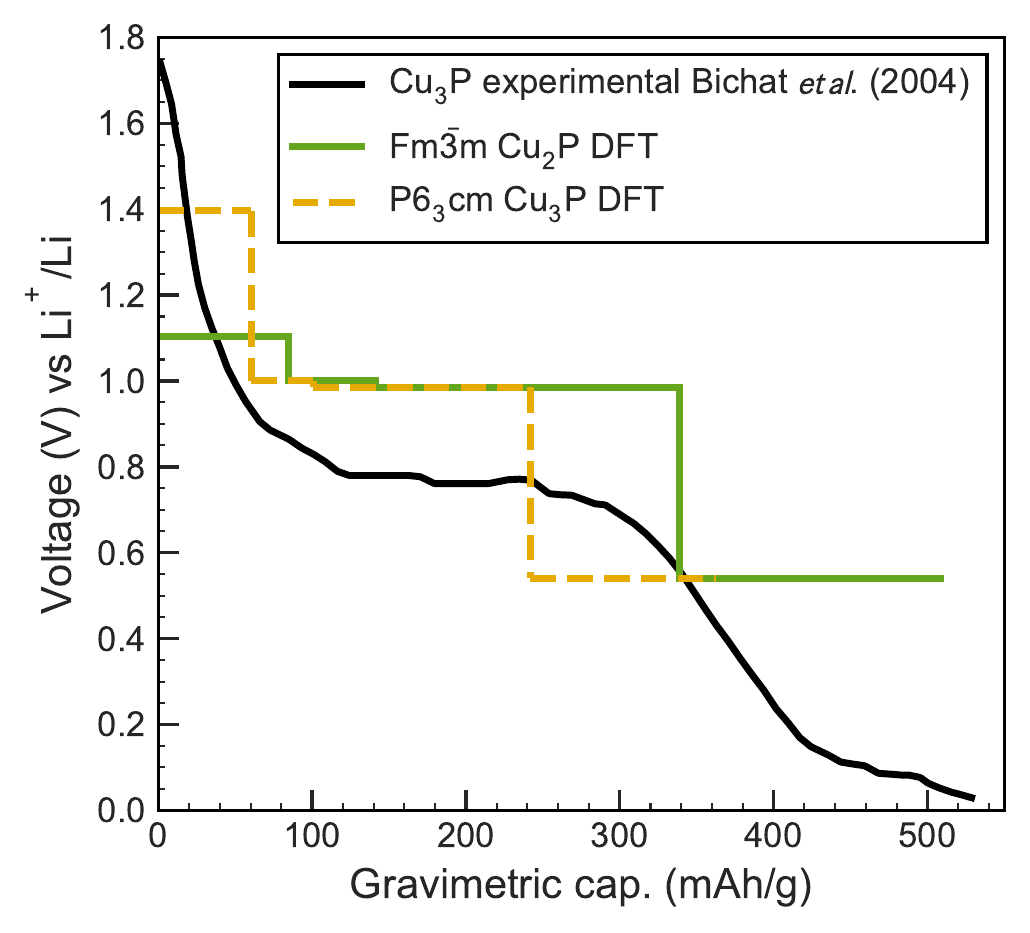}
  \caption{Ground-state voltage profile for $Fm\bar{3}m$--Cu$_2$P and $P6_3cm$--Cu$_3$P generated from the DFT ground-state structures of Cu--Li--P. The experimental profile was adapted from \cite{Bichat_2004}. The experimental onset voltage matches closely with Cu$_3$P and shows a similar capacity to Cu$_2$P, though this added capacity is likely due to oxide impurities in the experiment.}
  \label{fig:Cu2PCu3P_voltage}
\end{figure}

From this ternary hull, shown in Figure S9, the voltage profile shown in Figure \ref{fig:Cu2PCu3P_voltage} was constructed. This hull is calculated as usual, without incorporating vibrational effects at 0\,K. As the $P\bar{3}m1$--Cu$_2$LiP phase suggested in experiment \cite{Bichat_2004,SchlengerCu3P} is 39\,meV/atom above the convex hull, it cannot be in the voltage profile calculated in Figure \ref{fig:Cu2PCu3P_voltage}. The 0\,K voltage profile includes the $I4/mmm$--Cu$_2$LiP$_2$ phase, which has been previously synthesized through a solid state reaction \cite{schlenger1972kristallstrukturen} and is a high T$_c$ pnictide superconductor \cite{Cu2LiP2}. The $I4/mmm$--Cu$_2$LiP$_2$ phase has not, to our knowledge, been identified during cycling in Li-ion batteries previously.

Both Cu$_2$P and Cu$_3$P have the same overall reaction mechanism given by Equation \ref{eq:convrxn}, and the stable phases during the reaction from charging Cu$_3$P are given in Table \ref{tab:cunprxn}. These reactions show Cu$_2$P operates in a narrower voltage window than Cu$_3$P and has a higher predicted gravimetric capacity. The predicted structures forming at each capacity and voltage are given in the 4$^{th}$ column of Table \ref{tab:cunprxn}. Though $Fm\bar{3}m$--Cu$_2$P undergoes the same lithiation process as $P6_3cm$--Cu$_3$P, $Fm\bar{3}m$--Cu$_2$P has a higher capacity of 508\,mAh/g, and a higher average voltage of 0.86\,V versus Li/Li$^+$, while $P6_3cm$--Cu$_3$P has a capacity of 363\,mAh/g and average voltage of 0.91\,V.

Each plateau in the ground-state voltage profile in Figure \ref{fig:Cu2PCu3P_voltage} represents a three phase region of the ternary hull in which phases of Cu--Li--P are stable. Here, Cu$_3$P was `stabilized' on the ternary hull by artificially excluding the CuP$_2$--Cu$_2$P$_7$ and Cu$_2$P phases. This is an approximation of a convex hull in which Cu$_3$P is on the tie-line which does not effect the formation energy (and thus predicted voltages) of the other phases. The experimental voltage curve shown in Figure \ref{fig:Cu2PCu3P_voltage} from \cite{Bichat_2004} exhibits a similar trend in phase transitions along the cycle as the theoretical curve for Cu$_3$P.

\begin{table}[!htb]
    \centering
    \renewcommand{\arraystretch}{1.2}
    \begin{tabular}{|p{1.6cm}|p{4cm}|p{2.2cm}|p{5cm}|}
    \hline
    \multicolumn{1}{|c|}{Reaction} & \multicolumn{1}{|c|}{Gravimetric Capacity} & \multicolumn{1}{|c|}{Voltage (V)} & \multicolumn{1}{c|}{Reaction Pathway}\\
   \multicolumn{1}{|c|}{Stage} & \multicolumn{1}{|c|}{(mAh/g)}& & \\
    \hline
    \rowcolor{gray!30}\multicolumn{4}{|c|}{Cu$_2$P + 3 Li $\rightarrow$ Li$_3$P + 2 Cu}\\
    \hline
    I & 85 & 1.10 & $\frac{1}{2}$ Cu$_2$LiP$_2$ + Cu \\
    II & 141 & 1.00 &  $\frac{1}{6}$ Cu$_4$Li$_5$P$_6$ + $\frac{4}{3}$ Cu \\
    III & 338 & 0.98 & CuLi$_2$P + Cu \\
    IV & 509 & 0.54 & Li$_3$P + 2 Cu\\
    \hline
    \rowcolor{gray!30}\multicolumn{4}{|c|}{Cu$_3$P + 3 Li $\rightarrow$ Li$_3$P + 3 Cu}\\
    \hline
    I & 60 & 1.53 & $\frac{1}{2}$ Cu$_2$LiP$_2$ + 2 Cu  \\
    II & 101 & 1.00 &  $\frac{1}{6}$ Cu$_4$Li$_5$P$_6$ + $\frac{7}{3}$ Cu\\
    III & 242 & 0.98 & CuLi$_2$P + 2 Cu \\
    IV & 363 & 0.54 & Li$_3$P + 3 Cu \\ 
    \hline
    \end{tabular}
    \caption{Reaction pathways for Cu$_n$P + Li $\rightarrow$ Li$_3$P + $n$Cu}
    \label{tab:cunprxn}
\end{table}

\section{Conclusion}
\label{discussion}

Using 4 different computational crystal structure searching techniques on the copper phosphides, several structures which lie close to the convex hull (within 20\,meV/atom) were predicted, including $Cm$--Cu$_3$P$_{11}$, $I\bar{4}3d$--Cu$_3$P, and $Fm\bar{3}m$--Cu$_2$P; the experimentally characterized $P\bar{1}$--CuP$_{10}$, $C2/m$--Cu$_2$P$_7$ and $P2_1/c$--CuP$_2$ were all on the convex hull tie-line. By calculating the phonon dispersion curves of all structures within 20\,meV/atom of the Cu--P convex hull, we constructed a temperature dependent convex hull which predicted $Fm\bar{3}m$--Cu$_2$P to be stable up to 600\,K, while $I\bar{4}3d$--Cu$_3$P was destabilized with increasing temperature. We have also shown that the $Cmc2_1$--Cu$_8$P$_3$ phase formed from two Cu vacancies at the 6c Wyckoff positions of $P6_3cm$--Cu$_3$P is stabilized with increasing temperature, and is within 10\,meV/atom of the convex hull above 300\,K. Experimental diffractometry on single-crystals of Cu$_3$P suggests that the phase has a range of stoichiometries between Cu$_{2.6}$P and Cu$_{2.8}$P \cite{olofsson1972crystal,DeTrizio2015}, and Cu$_8$P$_3$ (or Cu$_{2.67}$P equivalently) is within these bounds.

In addition to confirming the stability of $Cmc2_1$--Cu$_8$P$_3$, we also confirmed that $Cm$--Cu$_3$P$_{11}$ remains metastable up to high temperatures as shown in the temperature dependent hull in Figure \ref{fig:CuP-thermo-hull}. While Cu$_3$P$_{11}$ is unlikely to be used as a Li-ion battery anode, given its high P content, and therefore susceptibility to volume expansion, it could be a novel phase to consider within the Cu--P phase diagram. CuP$_{10}$ was identified experimentally by preparing Cu$_2$P$_7$ in excess P \cite{bawohl2015phosphorus}; given the structural similarity between $Cm$--Cu$_3$P$_{11}$ and $C2/m$--Cu$_2$P$_7$ shown in Figure \ref{fig:Cu2P7Cu3P11struct}, it is possible that $Cm$--Cu$_3$P$_{11}$ is also formed in excess P. Using the PXRD patterns presented in Figure \ref{fig:Cu3P11-pdf}, it may be possible to distinguish the $Cm$--Cu$_3$P$_{11}$ phase from $C2/m$--Cu$_2$P$_7$ experimentally, from the change in peak intensity at 16$\degree$, and peak differences at 2$\theta$ values $< 20\degree$, however further experimental analysis is likely required given the low intensity of this peak. 

Finally, $Fm\bar{3}m$--Cu$_2$P is the only phase identified through crystal structure prediction which was found on the hull at 0\,K, and which remained on the convex hull at finite temperature, strongly suggesting it is possible to synthesize Cu$_2$P experimentally. Furthermore, its synthesis could provide a novel conversion anode, with favorable properties for Li-ion batteries. Hybrid functional calculations of the electronic properties of Cu$_2$P predict it to be isostructural and qualitatively similar electronically to both Rh$_2$P and Ir$_2$P, which are also $Fm\bar{3}m$ metals with dispersive bands at the Fermi level. This was confirmed using spin-polarized calculations, both with vector and scalar spin treatments, hybrid functional calculations using the HSE06 functional, and finally a projected band structure and density of states using PBE. This confirmation of the metallic nature of Cu$_2$P using a wide range of functionals and spin treatments suggests that this could be a better choice for anode than Li--P which are insulators with wide band gaps. Furthermore, the presence of such dispersive bands, suggest high electron mobility within the anode, which would mitigate fast charge transfer between the Cu current collector and Li-ions. Finally, given its higher capacity (509\,mAh/g) compared to Cu$_3$P, Cu$_2$P has potential as an experimentally realizable conversion anode which has a capacity that is competitive with graphite, conductive to promote electronic transfer within the anode, and less vulnerability to degradation compared to high P content conversion anodes due to reduced levels of cyclic volume changes.

\section{Acknowledgments}
The authors would like to thank Bartomeu Monserrat for his discussions on electronic structure, and Matthew Cliffe for his discussions on PXRD. We would also like to thank Can Ko\c cer for looking over the manuscript text. AFH acknowledges the financial support of the Gates Cambridge Trust and the Winton Programme for the Physics of Sustainability, University of Cambridge, UK. MLE acknowledges the Engineering and Physical Sciences Research Council (EPSRC) Centre for Doctoral Training in Computational Methods for Materials Science, UK, for funding (EP/L015552/1). AJM acknowledges funding from EPSRC (EP/P003532/1). The authors acknowledge networking support via the EPSRC Collaborative Computational Projects, CCP9 (EP/M022595/1) and CCP-NC (EP/T026642/1). This work was performed using resources provided by the Cambridge Service for Data Driven Discovery (CSD3) operated by the University of Cambridge Research Computing Service (www.csd3.cam.ac.uk), provided by Dell EMC and Intel using Tier-2 funding from the Engineering and Physical Sciences Research Council (capital grant EP/P020259/1), and DiRAC funding from the Science and Technology Facilities Council (www.dirac.ac.uk).

\section{Supporting Information Description}
The Supporting Information document contains details of the simulation of powder X-ray diffraction patterns and pair distribution functions, as well as the electronic band structure calculations using spin-orbit coupling. It contains all Supporting Information Figures referenced in the main text. All data is deposited at \href{https://doi.org/10.17863/CAM.52272}{https://doi.org/10.17863/CAM.52272} in the Cambridge Repository. All analysis is on Github at \href{https://github.com/harpaf13/data.copper-phosphides}{\texttt{harpaf13/data.copper-phosphides}} and \href{https://mybinder.org/v2/gh/harpaf13/data.copper-phosphides/master?filepath=CuP_results.ipynb}{Binder}.

\bibliography{apssamp}

\providecommand{\latin}[1]{#1}
\providecommand*\mcitethebibliography{\thebibliography}
\csname @ifundefined\endcsname{endmcitethebibliography}
  {\let\endmcitethebibliography\endthebibliography}{}
\begin{mcitethebibliography}{71}
\providecommand*\natexlab[1]{#1}
\providecommand*\mciteSetBstSublistMode[1]{}
\providecommand*\mciteSetBstMaxWidthForm[2]{}
\providecommand*\mciteBstWouldAddEndPuncttrue
  {\def\EndOfBibitem{\unskip.}}
\providecommand*\mciteBstWouldAddEndPunctfalse
  {\let\EndOfBibitem\relax}
\providecommand*\mciteSetBstMidEndSepPunct[3]{}
\providecommand*\mciteSetBstSublistLabelBeginEnd[3]{}
\providecommand*\EndOfBibitem{}
\mciteSetBstSublistMode{f}
\mciteSetBstMaxWidthForm{subitem}{(\alph{mcitesubitemcount})}
\mciteSetBstSublistLabelBeginEnd
  {\mcitemaxwidthsubitemform\space}
  {\relax}
  {\relax}

\bibitem[Ramireddy \latin{et~al.}(2015)Ramireddy, Xing, Rahman, Chen, Dutercq,
  Gunzelmann, and Glushenkov]{Ramireddy2015}
Ramireddy,~T.; Xing,~T.; Rahman,~M.~M.; Chen,~Y.; Dutercq,~Q.; Gunzelmann,~D.;
  Glushenkov,~A.~M. \emph{Journal of Materials Chemistry A} \textbf{2015},
  \emph{3}, 5572--5584\relax
\mciteBstWouldAddEndPuncttrue
\mciteSetBstMidEndSepPunct{\mcitedefaultmidpunct}
{\mcitedefaultendpunct}{\mcitedefaultseppunct}\relax
\EndOfBibitem
\bibitem[Bhatt and Lee(2019)Bhatt, and Lee]{BHATT201910852}
Bhatt,~M.~D.; Lee,~J.~Y. \emph{International Journal of Hydrogen Energy}
  \textbf{2019}, \emph{44}, 10852 -- 10905\relax
\mciteBstWouldAddEndPuncttrue
\mciteSetBstMidEndSepPunct{\mcitedefaultmidpunct}
{\mcitedefaultendpunct}{\mcitedefaultseppunct}\relax
\EndOfBibitem
\bibitem[Tarascon and Armand(2001)Tarascon, and Armand]{tarascon2011issues}
Tarascon,~J.-M.; Armand,~M. \emph{Nature} \textbf{2001}, \emph{414},
  359--367\relax
\mciteBstWouldAddEndPuncttrue
\mciteSetBstMidEndSepPunct{\mcitedefaultmidpunct}
{\mcitedefaultendpunct}{\mcitedefaultseppunct}\relax
\EndOfBibitem
\bibitem[Kirklin \latin{et~al.}(2013)Kirklin, Meredig, and
  Wolverton]{Wolverton2013}
Kirklin,~S.; Meredig,~B.; Wolverton,~C. \emph{Advanced Energy Materials}
  \textbf{2013}, \emph{3}, 252--262\relax
\mciteBstWouldAddEndPuncttrue
\mciteSetBstMidEndSepPunct{\mcitedefaultmidpunct}
{\mcitedefaultendpunct}{\mcitedefaultseppunct}\relax
\EndOfBibitem
\bibitem[Sun \latin{et~al.}(2016)Sun, Liu, Qu, and Li]{Sun2016}
Sun,~M.; Liu,~H.; Qu,~J.; Li,~J. \emph{Advanced Energy Materials}
  \textbf{2016}, \emph{6}, 1600087\relax
\mciteBstWouldAddEndPuncttrue
\mciteSetBstMidEndSepPunct{\mcitedefaultmidpunct}
{\mcitedefaultendpunct}{\mcitedefaultseppunct}\relax
\EndOfBibitem
\bibitem[Feng and Xue(2017)Feng, and Xue]{FengAdvances2017}
Feng,~L.; Xue,~H. \emph{Chem. Electro. Chem.} \textbf{2017}, \emph{4},
  20--34\relax
\mciteBstWouldAddEndPuncttrue
\mciteSetBstMidEndSepPunct{\mcitedefaultmidpunct}
{\mcitedefaultendpunct}{\mcitedefaultseppunct}\relax
\EndOfBibitem
\bibitem[Lu \latin{et~al.}(2018)Lu, Yu, and Lou]{LU2018972}
Lu,~Y.; Yu,~L.; Lou,~X. W.~D. \emph{Chem} \textbf{2018}, \emph{4}, 972 --
  996\relax
\mciteBstWouldAddEndPuncttrue
\mciteSetBstMidEndSepPunct{\mcitedefaultmidpunct}
{\mcitedefaultendpunct}{\mcitedefaultseppunct}\relax
\EndOfBibitem
\bibitem[Boyanov \latin{et~al.}(2006)Boyanov, Bernardi, Gillot, Dupont, Womes,
  Tarascon, Monconduit, and Doublet]{Boyanov2006}
Boyanov,~S.; Bernardi,~J.; Gillot,~F.; Dupont,~L.; Womes,~M.; Tarascon,~J.-M.;
  Monconduit,~L.; Doublet,~M.-L. \emph{Chemistry of Materials} \textbf{2006},
  \emph{18}, 3531--3538\relax
\mciteBstWouldAddEndPuncttrue
\mciteSetBstMidEndSepPunct{\mcitedefaultmidpunct}
{\mcitedefaultendpunct}{\mcitedefaultseppunct}\relax
\EndOfBibitem
\bibitem[Zhang \latin{et~al.}(2015)Zhang, Zhang, Feng, Liu, and
  Wang]{Zhang2015}
Zhang,~Y.; Zhang,~H.; Feng,~Y.; Liu,~L.; Wang,~Y. \emph{ACS Applied Materials
  \& Interfaces} \textbf{2015}, \emph{7}, 26684--26690\relax
\mciteBstWouldAddEndPuncttrue
\mciteSetBstMidEndSepPunct{\mcitedefaultmidpunct}
{\mcitedefaultendpunct}{\mcitedefaultseppunct}\relax
\EndOfBibitem
\bibitem[Lu \latin{et~al.}(2012)Lu, Wang, Mai, Xiang, Zhang, Li, Gu, Tu, and
  Mao]{Lu2012}
Lu,~Y.; Wang,~X.; Mai,~Y.; Xiang,~J.; Zhang,~H.; Li,~L.; Gu,~C.; Tu,~J.;
  Mao,~S.~X. \emph{The Journal of Physical Chemistry C} \textbf{2012},
  \emph{116}, 22217--22225\relax
\mciteBstWouldAddEndPuncttrue
\mciteSetBstMidEndSepPunct{\mcitedefaultmidpunct}
{\mcitedefaultendpunct}{\mcitedefaultseppunct}\relax
\EndOfBibitem
\bibitem[Kim and Manthiram(2017)Kim, and Manthiram]{Kim2017CuP2}
Kim,~S.-O.; Manthiram,~A. \emph{ACS Applied Materials \& Interfaces}
  \textbf{2017}, \emph{9}, 16221--16227\relax
\mciteBstWouldAddEndPuncttrue
\mciteSetBstMidEndSepPunct{\mcitedefaultmidpunct}
{\mcitedefaultendpunct}{\mcitedefaultseppunct}\relax
\EndOfBibitem
\bibitem[Wang \latin{et~al.}(2003)Wang, Yang, Xie, Wang, and Wen]{WANG2003480}
Wang,~K.; Yang,~J.; Xie,~J.; Wang,~B.; Wen,~Z. \emph{Electrochemistry
  Communications} \textbf{2003}, \emph{5}, 480 -- 483\relax
\mciteBstWouldAddEndPuncttrue
\mciteSetBstMidEndSepPunct{\mcitedefaultmidpunct}
{\mcitedefaultendpunct}{\mcitedefaultseppunct}\relax
\EndOfBibitem
\bibitem[Bichat \latin{et~al.}(2004)Bichat, Politova, Pfeiffer, Tancret,
  Monconduit, Pascal, Brousse, and Favier]{BICHAT200480}
Bichat,~M.-P.; Politova,~T.; Pfeiffer,~H.; Tancret,~F.; Monconduit,~L.;
  Pascal,~J.-L.; Brousse,~T.; Favier,~F. \emph{Journal of Power Sources}
  \textbf{2004}, \emph{136}, 80 -- 87\relax
\mciteBstWouldAddEndPuncttrue
\mciteSetBstMidEndSepPunct{\mcitedefaultmidpunct}
{\mcitedefaultendpunct}{\mcitedefaultseppunct}\relax
\EndOfBibitem
\bibitem[Pfeiffer \latin{et~al.}(2004)Pfeiffer, Tancret, Bichat, Monconduit,
  Favier, and Brousse]{PFEIFFER2004263}
Pfeiffer,~H.; Tancret,~F.; Bichat,~M.-P.; Monconduit,~L.; Favier,~F.;
  Brousse,~T. \emph{Electrochemistry Communications} \textbf{2004}, \emph{6},
  263 -- 267\relax
\mciteBstWouldAddEndPuncttrue
\mciteSetBstMidEndSepPunct{\mcitedefaultmidpunct}
{\mcitedefaultendpunct}{\mcitedefaultseppunct}\relax
\EndOfBibitem
\bibitem[Villevieille \latin{et~al.}(2008)Villevieille, Robert, Taberna, Bazin,
  Simon, and Monconduit]{villevieille2008good}
Villevieille,~C.; Robert,~F.; Taberna,~P.~L.; Bazin,~L.; Simon,~P.;
  Monconduit,~L. \emph{Journal of Materials Chemistry} \textbf{2008},
  \emph{18}, 5956--5960\relax
\mciteBstWouldAddEndPuncttrue
\mciteSetBstMidEndSepPunct{\mcitedefaultmidpunct}
{\mcitedefaultendpunct}{\mcitedefaultseppunct}\relax
\EndOfBibitem
\bibitem[Ni \latin{et~al.}(2014)Ni, Ma, Lv, Yang, and Zhang]{Ni2014Cu3P}
Ni,~S.; Ma,~J.; Lv,~X.; Yang,~X.; Zhang,~L. \emph{Journal of Materials
  Chemistry A} \textbf{2014}, \emph{2}, 20506--20509\relax
\mciteBstWouldAddEndPuncttrue
\mciteSetBstMidEndSepPunct{\mcitedefaultmidpunct}
{\mcitedefaultendpunct}{\mcitedefaultseppunct}\relax
\EndOfBibitem
\bibitem[Chandrasekar and Mitra(2013)Chandrasekar, and
  Mitra]{chandrasekar2013thin}
Chandrasekar,~M.; Mitra,~S. \emph{Electrochimica Acta} \textbf{2013},
  \emph{92}, 47--54\relax
\mciteBstWouldAddEndPuncttrue
\mciteSetBstMidEndSepPunct{\mcitedefaultmidpunct}
{\mcitedefaultendpunct}{\mcitedefaultseppunct}\relax
\EndOfBibitem
\bibitem[Curtarolo \latin{et~al.}(2012)Curtarolo, Setyawan, Hart, Jahnatek,
  Chepulskii, Taylor, Wang, Xue, Yang, Levy, Mehl, Stokes, Demchenko, and
  Morgan]{CURTAROLO2012218}
Curtarolo,~S.; Setyawan,~W.; Hart,~G.~L.; Jahnatek,~M.; Chepulskii,~R.~V.;
  Taylor,~R.~H.; Wang,~S.; Xue,~J.; Yang,~K.; Levy,~O.; Mehl,~M.~J.;
  Stokes,~H.~T.; Demchenko,~D.~O.; Morgan,~D. \emph{Computational Materials
  Science} \textbf{2012}, \emph{58}, 218 -- 226\relax
\mciteBstWouldAddEndPuncttrue
\mciteSetBstMidEndSepPunct{\mcitedefaultmidpunct}
{\mcitedefaultendpunct}{\mcitedefaultseppunct}\relax
\EndOfBibitem
\bibitem[Saal \latin{et~al.}(2013)Saal, Kirklin, Aykol, Meredig, and
  Wolverton]{saal2013materials}
Saal,~J.~E.; Kirklin,~S.; Aykol,~M.; Meredig,~B.; Wolverton,~C. \emph{JOM}
  \textbf{2013}, \emph{65}, 1501--1509\relax
\mciteBstWouldAddEndPuncttrue
\mciteSetBstMidEndSepPunct{\mcitedefaultmidpunct}
{\mcitedefaultendpunct}{\mcitedefaultseppunct}\relax
\EndOfBibitem
\bibitem[Hautier \latin{et~al.}(2011)Hautier, Fischer, Ehrlacher, Jain, and
  Ceder]{hautier2011data}
Hautier,~G.; Fischer,~C.; Ehrlacher,~V.; Jain,~A.; Ceder,~G. \emph{Inorganic
  Chemistry} \textbf{2011}, \emph{50}, 656--663\relax
\mciteBstWouldAddEndPuncttrue
\mciteSetBstMidEndSepPunct{\mcitedefaultmidpunct}
{\mcitedefaultendpunct}{\mcitedefaultseppunct}\relax
\EndOfBibitem
\bibitem[Evans(2016)]{matador}
Evans,~M. \emph{GitHub} \textbf{2016},
  \url{https://github.com/ml-evs/matador/releases/tag/0.9} (accessed May 11,
  2020)\relax
\mciteBstWouldAddEndPuncttrue
\mciteSetBstMidEndSepPunct{\mcitedefaultmidpunct}
{\mcitedefaultendpunct}{\mcitedefaultseppunct}\relax
\EndOfBibitem
\bibitem[Evans(2017)]{ilustrado}
Evans,~M. \emph{GitHub} \textbf{2017},
  \url{https://github.com/ml-evs/ilustrado/releases/tag/0.3b} (accessed May 11,
  2020)\relax
\mciteBstWouldAddEndPuncttrue
\mciteSetBstMidEndSepPunct{\mcitedefaultmidpunct}
{\mcitedefaultendpunct}{\mcitedefaultseppunct}\relax
\EndOfBibitem
\bibitem[Harper \latin{et~al.}(2019)Harper, Evans, Darby, Karasulu,
  Ko{\c{c}}er, Nelson, and Morris]{harper2019ab}
Harper,~A.~F.; Evans,~M.~L.; Darby,~J.~P.; Karasulu,~B.; Ko{\c{c}}er,~C.~P.;
  Nelson,~J.~R.; Morris,~A.~J. \emph{Johnson Matthey Technology Review}
  \textbf{2019}, \emph{64}, 103--118\relax
\mciteBstWouldAddEndPuncttrue
\mciteSetBstMidEndSepPunct{\mcitedefaultmidpunct}
{\mcitedefaultendpunct}{\mcitedefaultseppunct}\relax
\EndOfBibitem
\bibitem[See \latin{et~al.}(2014)See, Leskes, Griffin, Britto, Matthews, Emly,
  Van~der Ven, Wright, Morris, Grey, \latin{et~al.} others]{see2014ab}
others,, \latin{et~al.}  \emph{Journal of the American Chemical Society}
  \textbf{2014}, \emph{136}, 16368--16377\relax
\mciteBstWouldAddEndPuncttrue
\mciteSetBstMidEndSepPunct{\mcitedefaultmidpunct}
{\mcitedefaultendpunct}{\mcitedefaultseppunct}\relax
\EndOfBibitem
\bibitem[Mayo and Morris(2017)Mayo, and Morris]{mayo2017structure}
Mayo,~M.; Morris,~A.~J. \emph{Chemistry of Materials} \textbf{2017}, \emph{29},
  5787--5795\relax
\mciteBstWouldAddEndPuncttrue
\mciteSetBstMidEndSepPunct{\mcitedefaultmidpunct}
{\mcitedefaultendpunct}{\mcitedefaultseppunct}\relax
\EndOfBibitem
\bibitem[Mayo \latin{et~al.}(2016)Mayo, Griffith, Pickard, and
  Morris]{mayo2016ab}
Mayo,~M.; Griffith,~K.~J.; Pickard,~C.~J.; Morris,~A.~J. \emph{Chemistry of
  Materials} \textbf{2016}, \emph{28}, 2011--2021\relax
\mciteBstWouldAddEndPuncttrue
\mciteSetBstMidEndSepPunct{\mcitedefaultmidpunct}
{\mcitedefaultendpunct}{\mcitedefaultseppunct}\relax
\EndOfBibitem
\bibitem[Marbella \latin{et~al.}(2018)Marbella, Evans, Groh, Nelson, Griffith,
  Morris, and Grey]{MarbellaACS2018}
Marbella,~L.~E.; Evans,~M.~L.; Groh,~M.~F.; Nelson,~J.; Griffith,~K.~J.;
  Morris,~A.~J.; Grey,~C.~P. \emph{Journal of the American Chemical Society}
  \textbf{2018}, \emph{140}, 7994--8004\relax
\mciteBstWouldAddEndPuncttrue
\mciteSetBstMidEndSepPunct{\mcitedefaultmidpunct}
{\mcitedefaultendpunct}{\mcitedefaultseppunct}\relax
\EndOfBibitem
\bibitem[Hellenbrandt(2004)]{hellenbrandt2004inorganic}
Hellenbrandt,~M. \emph{Crystallography Reviews} \textbf{2004}, \emph{10},
  17--22\relax
\mciteBstWouldAddEndPuncttrue
\mciteSetBstMidEndSepPunct{\mcitedefaultmidpunct}
{\mcitedefaultendpunct}{\mcitedefaultseppunct}\relax
\EndOfBibitem
\bibitem[Pickard and Needs(2011)Pickard, and Needs]{pickard2011ab}
Pickard,~C.~J.; Needs,~R. \emph{Journal of Physics: Condensed Matter}
  \textbf{2011}, \emph{23}, 053201\relax
\mciteBstWouldAddEndPuncttrue
\mciteSetBstMidEndSepPunct{\mcitedefaultmidpunct}
{\mcitedefaultendpunct}{\mcitedefaultseppunct}\relax
\EndOfBibitem
\bibitem[Deaven and Ho(1995)Deaven, and Ho]{Deaven1995}
Deaven,~D.~M.; Ho,~K.~M. \emph{Physical Review Letters} \textbf{1995},
  \emph{75}, 288--291\relax
\mciteBstWouldAddEndPuncttrue
\mciteSetBstMidEndSepPunct{\mcitedefaultmidpunct}
{\mcitedefaultendpunct}{\mcitedefaultseppunct}\relax
\EndOfBibitem
\bibitem[Clark \latin{et~al.}(2005)Clark, Segall, Pickard, Hasnip, Probert,
  Refson, and Payne]{clark2005first}
Clark,~S.~J.; Segall,~M.~D.; Pickard,~C.~J.; Hasnip,~P.~J.; Probert,~M.~I.;
  Refson,~K.; Payne,~M.~C. \emph{Zeitschrift f{\"u}r
  Kristallographie-Crystalline Materials} \textbf{2005}, \emph{220},
  567--570\relax
\mciteBstWouldAddEndPuncttrue
\mciteSetBstMidEndSepPunct{\mcitedefaultmidpunct}
{\mcitedefaultendpunct}{\mcitedefaultseppunct}\relax
\EndOfBibitem
\bibitem[Perdew \latin{et~al.}(1996)Perdew, Burke, and
  Ernzerhof]{perdew1996generalized}
Perdew,~J.~P.; Burke,~K.; Ernzerhof,~M. \emph{Physical Review Letters}
  \textbf{1996}, \emph{77}, 3865\relax
\mciteBstWouldAddEndPuncttrue
\mciteSetBstMidEndSepPunct{\mcitedefaultmidpunct}
{\mcitedefaultendpunct}{\mcitedefaultseppunct}\relax
\EndOfBibitem
\bibitem[Vanderbilt(1990)]{vanderbilt1990soft}
Vanderbilt,~D. \emph{Physical Review B} \textbf{1990}, \emph{41}, 7892\relax
\mciteBstWouldAddEndPuncttrue
\mciteSetBstMidEndSepPunct{\mcitedefaultmidpunct}
{\mcitedefaultendpunct}{\mcitedefaultseppunct}\relax
\EndOfBibitem
\bibitem[Sun \latin{et~al.}(2012)Sun, Li, Sun, Yu, Wang, and
  Xie]{sun2012electrochemical}
Sun,~L.-Q.; Li,~M.-J.; Sun,~K.; Yu,~S.-H.; Wang,~R.-S.; Xie,~H.-M. \emph{The
  Journal of Physical Chemistry C} \textbf{2012}, \emph{116},
  14772--14779\relax
\mciteBstWouldAddEndPuncttrue
\mciteSetBstMidEndSepPunct{\mcitedefaultmidpunct}
{\mcitedefaultendpunct}{\mcitedefaultseppunct}\relax
\EndOfBibitem
\bibitem[Urban \latin{et~al.}(2016)Urban, Seo, and
  Ceder]{urban2016computational}
Urban,~A.; Seo,~D.-H.; Ceder,~G. \emph{npj Computational Materials}
  \textbf{2016}, \emph{2}, 1--13\relax
\mciteBstWouldAddEndPuncttrue
\mciteSetBstMidEndSepPunct{\mcitedefaultmidpunct}
{\mcitedefaultendpunct}{\mcitedefaultseppunct}\relax
\EndOfBibitem
\bibitem[Hinuma \latin{et~al.}(2017)Hinuma, Pizzi, Kumagai, Oba, and
  Tanaka]{HINUMA2017140}
Hinuma,~Y.; Pizzi,~G.; Kumagai,~Y.; Oba,~F.; Tanaka,~I. \emph{Computational
  Materials Science} \textbf{2017}, \emph{128}, 140 -- 184\relax
\mciteBstWouldAddEndPuncttrue
\mciteSetBstMidEndSepPunct{\mcitedefaultmidpunct}
{\mcitedefaultendpunct}{\mcitedefaultseppunct}\relax
\EndOfBibitem
\bibitem[Togo and Tanaka(2018)Togo, and Tanaka]{togo_spglib_2018}
Togo,~A.; Tanaka,~I. \emph{arXiv} \textbf{2018},
  \url{http://arxiv.org/abs/1808.01590} (accessed Feb 19, 2020)\relax
\mciteBstWouldAddEndPuncttrue
\mciteSetBstMidEndSepPunct{\mcitedefaultmidpunct}
{\mcitedefaultendpunct}{\mcitedefaultseppunct}\relax
\EndOfBibitem
\bibitem[Morris \latin{et~al.}(2014)Morris, Nicholls, Pickard, and
  Yates]{morris2014optados}
Morris,~A.~J.; Nicholls,~R.~J.; Pickard,~C.~J.; Yates,~J.~R. \emph{Computer
  Physics Communications} \textbf{2014}, \emph{185}, 1477--1485\relax
\mciteBstWouldAddEndPuncttrue
\mciteSetBstMidEndSepPunct{\mcitedefaultmidpunct}
{\mcitedefaultendpunct}{\mcitedefaultseppunct}\relax
\EndOfBibitem
\bibitem[{Nicholls} \latin{et~al.}(2012){Nicholls}, {Morris}, {Pickard}, and
  {Yates}]{NichollsOptaDOS}
{Nicholls},~R.~J.; {Morris},~A.~J.; {Pickard},~C.~J.; {Yates},~J.~R. {OptaDOS -
  a new tool for EELS calculations}. Journal of Physics Conference Series.
  2012; p 012062\relax
\mciteBstWouldAddEndPuncttrue
\mciteSetBstMidEndSepPunct{\mcitedefaultmidpunct}
{\mcitedefaultendpunct}{\mcitedefaultseppunct}\relax
\EndOfBibitem
\bibitem[Tkatchenko \latin{et~al.}(2012)Tkatchenko, DiStasio~Jr, Car, and
  Scheffler]{tkatchenko2012accurate}
Tkatchenko,~A.; DiStasio~Jr,~R.~A.; Car,~R.; Scheffler,~M. \emph{Physical
  Review Letters} \textbf{2012}, \emph{108}, 236402\relax
\mciteBstWouldAddEndPuncttrue
\mciteSetBstMidEndSepPunct{\mcitedefaultmidpunct}
{\mcitedefaultendpunct}{\mcitedefaultseppunct}\relax
\EndOfBibitem
\bibitem[Sun \latin{et~al.}(2016)Sun, Dacek, Ong, Hautier, Jain, Richards,
  Gamst, Persson, and Ceder]{sun2016thermodynamic}
Sun,~W.; Dacek,~S.~T.; Ong,~S.~P.; Hautier,~G.; Jain,~A.; Richards,~W.~D.;
  Gamst,~A.~C.; Persson,~K.~A.; Ceder,~G. \emph{Science Advances}
  \textbf{2016}, \emph{2}, e1600225\relax
\mciteBstWouldAddEndPuncttrue
\mciteSetBstMidEndSepPunct{\mcitedefaultmidpunct}
{\mcitedefaultendpunct}{\mcitedefaultseppunct}\relax
\EndOfBibitem
\bibitem[Zhang \latin{et~al.}(2018)Zhang, Reilly, Tkatchenko, and
  Scheffler]{zhang2018performance}
Zhang,~G.-X.; Reilly,~A.~M.; Tkatchenko,~A.; Scheffler,~M. \emph{New Journal of
  Physics} \textbf{2018}, \emph{20}, 063020\relax
\mciteBstWouldAddEndPuncttrue
\mciteSetBstMidEndSepPunct{\mcitedefaultmidpunct}
{\mcitedefaultendpunct}{\mcitedefaultseppunct}\relax
\EndOfBibitem
\bibitem[Olofsson(1972)]{olofsson1972crystal}
Olofsson,~O. \emph{Acta Chemica Scandinavica} \textbf{1972}, \emph{26},
  2777--2787\relax
\mciteBstWouldAddEndPuncttrue
\mciteSetBstMidEndSepPunct{\mcitedefaultmidpunct}
{\mcitedefaultendpunct}{\mcitedefaultseppunct}\relax
\EndOfBibitem
\bibitem[Lange \latin{et~al.}(2008)Lange, Bawohl, Weihrich, and Nilges]{CuP10}
Lange,~S.; Bawohl,~M.; Weihrich,~R.; Nilges,~T. \emph{Angewandte Chemie
  International Edition} \textbf{2008}, \emph{47}, 5654--5657\relax
\mciteBstWouldAddEndPuncttrue
\mciteSetBstMidEndSepPunct{\mcitedefaultmidpunct}
{\mcitedefaultendpunct}{\mcitedefaultseppunct}\relax
\EndOfBibitem
\bibitem[M{\o}uller and Jeitschko(1982)M{\o}uller, and
  Jeitschko]{mouller1982darstellung}
M{\o}uller,~M.; Jeitschko,~W. \emph{Journal for Inorganic and General
  Chemistry} \textbf{1982}, \emph{491}, 225--236\relax
\mciteBstWouldAddEndPuncttrue
\mciteSetBstMidEndSepPunct{\mcitedefaultmidpunct}
{\mcitedefaultendpunct}{\mcitedefaultseppunct}\relax
\EndOfBibitem
\bibitem[Schlenger \latin{et~al.}(1971)Schlenger, Jacobs, and
  Juza]{SchlengerCu3P}
Schlenger,~H.; Jacobs,~H.; Juza,~R. \emph{Journal of Inorganic and General
  Chemistry} \textbf{1971}, \emph{385}, 177--201\relax
\mciteBstWouldAddEndPuncttrue
\mciteSetBstMidEndSepPunct{\mcitedefaultmidpunct}
{\mcitedefaultendpunct}{\mcitedefaultseppunct}\relax
\EndOfBibitem
\bibitem[De~Trizio \latin{et~al.}(2015)De~Trizio, Gaspari, Bertoni, Kriegel,
  Moretti, Scotognella, Maserati, Zhang, Messina, Prato, \latin{et~al.}
  others]{DeTrizio2015}
others,, \latin{et~al.}  \emph{Chemistry of Materials} \textbf{2015},
  \emph{27}, 1120--1128\relax
\mciteBstWouldAddEndPuncttrue
\mciteSetBstMidEndSepPunct{\mcitedefaultmidpunct}
{\mcitedefaultendpunct}{\mcitedefaultseppunct}\relax
\EndOfBibitem
\bibitem[Otte(1961)]{otte1961lattice}
Otte,~H.~M. \emph{Journal of Applied Physics} \textbf{1961}, \emph{32},
  1536--1546\relax
\mciteBstWouldAddEndPuncttrue
\mciteSetBstMidEndSepPunct{\mcitedefaultmidpunct}
{\mcitedefaultendpunct}{\mcitedefaultseppunct}\relax
\EndOfBibitem
\bibitem[Steenberg(1938)]{steenberg1938crystal}
Steenberg,~B. \emph{Arkiv Foer Kemi, Meralogi Och Geologi a Argka}
  \textbf{1938}, \emph{12}\relax
\mciteBstWouldAddEndPuncttrue
\mciteSetBstMidEndSepPunct{\mcitedefaultmidpunct}
{\mcitedefaultendpunct}{\mcitedefaultseppunct}\relax
\EndOfBibitem
\bibitem[Owusu \latin{et~al.}(1972)Owusu, Jawad, Lundström, and
  Rundqvist]{Owusu_1972}
Owusu,~M.; Jawad,~H.; Lundström,~T.; Rundqvist,~S. \emph{Physica Scripta}
  \textbf{1972}, \emph{6}, 67--70\relax
\mciteBstWouldAddEndPuncttrue
\mciteSetBstMidEndSepPunct{\mcitedefaultmidpunct}
{\mcitedefaultendpunct}{\mcitedefaultseppunct}\relax
\EndOfBibitem
\bibitem[Zumbusch(1940)]{zumbusch1940structures}
Zumbusch,~M. \emph{Journal for Inorganic and General Chemistry} \textbf{1940},
  \emph{243}, 322--329\relax
\mciteBstWouldAddEndPuncttrue
\mciteSetBstMidEndSepPunct{\mcitedefaultmidpunct}
{\mcitedefaultendpunct}{\mcitedefaultseppunct}\relax
\EndOfBibitem
\bibitem[Moeller and Jeitschko(1981)Moeller, and Jeitschko]{MoellerInorg1981}
Moeller,~M.~H.; Jeitschko,~W. \emph{Inorganic Chemistry} \textbf{1981},
  \emph{20}, 828--833\relax
\mciteBstWouldAddEndPuncttrue
\mciteSetBstMidEndSepPunct{\mcitedefaultmidpunct}
{\mcitedefaultendpunct}{\mcitedefaultseppunct}\relax
\EndOfBibitem
\bibitem[Crichton \latin{et~al.}(2003)Crichton, Mezouar, Monaco, and
  Falconi]{crichton2003phosphorus}
Crichton,~W.~A.; Mezouar,~M.; Monaco,~G.; Falconi,~S. \emph{Powder Diffraction}
  \textbf{2003}, \emph{18}, 155--158\relax
\mciteBstWouldAddEndPuncttrue
\mciteSetBstMidEndSepPunct{\mcitedefaultmidpunct}
{\mcitedefaultendpunct}{\mcitedefaultseppunct}\relax
\EndOfBibitem
\bibitem[Thurn and Krebs(1969)Thurn, and Krebs]{Thurna06597}
Thurn,~H.; Krebs,~H. \emph{Acta Crystallographica Section B: Structural
  Crystallography and Crystal Chemistry} \textbf{1969}, \emph{25},
  125--135\relax
\mciteBstWouldAddEndPuncttrue
\mciteSetBstMidEndSepPunct{\mcitedefaultmidpunct}
{\mcitedefaultendpunct}{\mcitedefaultseppunct}\relax
\EndOfBibitem
\bibitem[Wolff \latin{et~al.}(2018)Wolff, Doert, Hunger, Kaiser, Pallmann,
  Reinhold, Yogendra, Giebeler, Sichelschmidt, Schnelle, \latin{et~al.}
  others]{wolff2018low}
others,, \latin{et~al.}  \emph{Chemistry of Materials} \textbf{2018},
  \emph{30}, 7111--7123\relax
\mciteBstWouldAddEndPuncttrue
\mciteSetBstMidEndSepPunct{\mcitedefaultmidpunct}
{\mcitedefaultendpunct}{\mcitedefaultseppunct}\relax
\EndOfBibitem
\bibitem[Yang and Ganz(2016)Yang, and Ganz]{C6CP01860B}
Yang,~L.-M.; Ganz,~E. \emph{Physical Chemistry Chemical Physics} \textbf{2016},
  \emph{18}, 17586--17591\relax
\mciteBstWouldAddEndPuncttrue
\mciteSetBstMidEndSepPunct{\mcitedefaultmidpunct}
{\mcitedefaultendpunct}{\mcitedefaultseppunct}\relax
\EndOfBibitem
\bibitem[Iglesias and Nowacki(1977)Iglesias, and
  Nowacki]{iglesias1977refinement}
Iglesias,~J.; Nowacki,~W. \emph{Zeitschrift f{\"u}r
  Kristallographie-Crystalline Materials} \textbf{1977}, \emph{145},
  334--345\relax
\mciteBstWouldAddEndPuncttrue
\mciteSetBstMidEndSepPunct{\mcitedefaultmidpunct}
{\mcitedefaultendpunct}{\mcitedefaultseppunct}\relax
\EndOfBibitem
\bibitem[Rundqvist(1960)]{rundqvist1960phosphides}
Rundqvist,~S. \emph{Nature} \textbf{1960}, \emph{185}, 31--32\relax
\mciteBstWouldAddEndPuncttrue
\mciteSetBstMidEndSepPunct{\mcitedefaultmidpunct}
{\mcitedefaultendpunct}{\mcitedefaultseppunct}\relax
\EndOfBibitem
\bibitem[Jain \latin{et~al.}(2013)Jain, Ong, Hautier, Chen, Richards, Dacek,
  Cholia, Gunter, Skinner, Ceder, and Persson]{Jain2013}
Jain,~A.; Ong,~S.~P.; Hautier,~G.; Chen,~W.; Richards,~W.~D.; Dacek,~S.;
  Cholia,~S.; Gunter,~D.; Skinner,~D.; Ceder,~G.; Persson,~K.~A. \emph{APL
  Materials} \textbf{2013}, \emph{1}, 011002\relax
\mciteBstWouldAddEndPuncttrue
\mciteSetBstMidEndSepPunct{\mcitedefaultmidpunct}
{\mcitedefaultendpunct}{\mcitedefaultseppunct}\relax
\EndOfBibitem
\bibitem[Larsson(1965)]{larsson1965x}
Larsson,~E. \emph{Ark. Kemi} \textbf{1965}, \emph{23}, 335--365\relax
\mciteBstWouldAddEndPuncttrue
\mciteSetBstMidEndSepPunct{\mcitedefaultmidpunct}
{\mcitedefaultendpunct}{\mcitedefaultseppunct}\relax
\EndOfBibitem
\bibitem[Carlsson \latin{et~al.}(1973)Carlsson, G{\"o}lin, and
  Rundqvist]{carlsson1973determination}
Carlsson,~B.; G{\"o}lin,~M.; Rundqvist,~S. \emph{Journal of Solid State
  Chemistry} \textbf{1973}, \emph{8}, 57--67\relax
\mciteBstWouldAddEndPuncttrue
\mciteSetBstMidEndSepPunct{\mcitedefaultmidpunct}
{\mcitedefaultendpunct}{\mcitedefaultseppunct}\relax
\EndOfBibitem
\bibitem[Song \latin{et~al.}(2009)Song, Kang, Kim, Park, and
  Kwon]{song2009nature}
Song,~M.-S.; Kang,~Y.-M.; Kim,~Y.-I.; Park,~K.-S.; Kwon,~H.-S. \emph{Inorganic
  Chemistry} \textbf{2009}, \emph{48}, 8271--8275\relax
\mciteBstWouldAddEndPuncttrue
\mciteSetBstMidEndSepPunct{\mcitedefaultmidpunct}
{\mcitedefaultendpunct}{\mcitedefaultseppunct}\relax
\EndOfBibitem
\bibitem[Baroni \latin{et~al.}(2001)Baroni, de~Gironcoli, Dal~Corso, and
  Giannozzi]{Baroni2001}
Baroni,~S.; de~Gironcoli,~S.; Dal~Corso,~A.; Giannozzi,~P. \emph{Rev. Mod.
  Phys.} \textbf{2001}, \emph{73}, 515--562\relax
\mciteBstWouldAddEndPuncttrue
\mciteSetBstMidEndSepPunct{\mcitedefaultmidpunct}
{\mcitedefaultendpunct}{\mcitedefaultseppunct}\relax
\EndOfBibitem
\bibitem[Van~de Walle \latin{et~al.}(2004)Van~de Walle, Moser, and
  Gasior]{van2004first}
Van~de Walle,~A.; Moser,~Z.; Gasior,~W. \emph{Archives of Metallurgy}
  \textbf{2004}, \emph{49}, 535--544\relax
\mciteBstWouldAddEndPuncttrue
\mciteSetBstMidEndSepPunct{\mcitedefaultmidpunct}
{\mcitedefaultendpunct}{\mcitedefaultseppunct}\relax
\EndOfBibitem
\bibitem[Okamoto(2011)]{okamoto2011cu}
Okamoto,~H. \emph{Journal of Phase Equilibria and Diffusion} \textbf{2011},
  \emph{32}, 172--172\relax
\mciteBstWouldAddEndPuncttrue
\mciteSetBstMidEndSepPunct{\mcitedefaultmidpunct}
{\mcitedefaultendpunct}{\mcitedefaultseppunct}\relax
\EndOfBibitem
\bibitem[Fang \latin{et~al.}(2019)Fang, Li, Zhang, Zhang, Yang, Lee, Lee,
  Alvarado, Schroeder, Yang, \latin{et~al.} others]{fang2019quantifying}
others,, \latin{et~al.}  \emph{Nature} \textbf{2019}, \emph{572},
  511--515\relax
\mciteBstWouldAddEndPuncttrue
\mciteSetBstMidEndSepPunct{\mcitedefaultmidpunct}
{\mcitedefaultendpunct}{\mcitedefaultseppunct}\relax
\EndOfBibitem
\bibitem[Bichat \latin{et~al.}(2004)Bichat, Politova, Pascal, Favier, and
  Monconduit]{Bichat_2004}
Bichat,~M.~P.; Politova,~T.; Pascal,~J.~L.; Favier,~F.; Monconduit,~L.
  \emph{Journal of The Electrochemical Society} \textbf{2004}, \emph{151},
  A2074\relax
\mciteBstWouldAddEndPuncttrue
\mciteSetBstMidEndSepPunct{\mcitedefaultmidpunct}
{\mcitedefaultendpunct}{\mcitedefaultseppunct}\relax
\EndOfBibitem
\bibitem[Schlenger and Jacobs(1972)Schlenger, and
  Jacobs]{schlenger1972kristallstrukturen}
Schlenger,~H.; Jacobs,~H. \emph{Acta Crystallographica Section B: Structural
  Crystallography and Crystal Chemistry} \textbf{1972}, \emph{28},
  327--327\relax
\mciteBstWouldAddEndPuncttrue
\mciteSetBstMidEndSepPunct{\mcitedefaultmidpunct}
{\mcitedefaultendpunct}{\mcitedefaultseppunct}\relax
\EndOfBibitem
\bibitem[Han \latin{et~al.}(2010)Han, Zhou, Cheng, and Goodenough]{Cu2LiP2}
Han,~J.-T.; Zhou,~J.-S.; Cheng,~J.-G.; Goodenough,~J.~B. \emph{Journal of the
  American Chemical Society} \textbf{2010}, \emph{132}, 908--909\relax
\mciteBstWouldAddEndPuncttrue
\mciteSetBstMidEndSepPunct{\mcitedefaultmidpunct}
{\mcitedefaultendpunct}{\mcitedefaultseppunct}\relax
\EndOfBibitem
\bibitem[Bawohl and Nilges(2015)Bawohl, and Nilges]{bawohl2015phosphorus}
Bawohl,~M.; Nilges,~T. \emph{{Zeitschrift f{\"u}r Anorganische und Allgemeine
  Chemie}} \textbf{2015}, \emph{641}, 304--310\relax
\mciteBstWouldAddEndPuncttrue
\mciteSetBstMidEndSepPunct{\mcitedefaultmidpunct}
{\mcitedefaultendpunct}{\mcitedefaultseppunct}\relax
\EndOfBibitem
\end{mcitethebibliography}

\pagebreak
\section{For Table of Contents Only}
\begin{figure}
    \centering
    \includegraphics{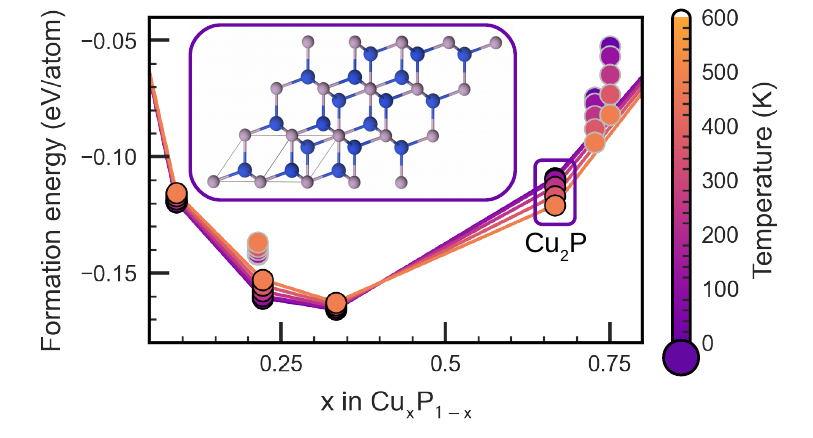}
\end{figure}

\pagebreak
\setcounter{equation}{0}
\setcounter{figure}{0}
\setcounter{table}{0}
\setcounter{page}{1}
\makeatletter
\renewcommand{\theequation}{S\arabic{equation}}
\renewcommand{\thefigure}{S\arabic{figure}}
\section{Supporting Information}

 \graphicspath{ {./supp_figs/} }
\subsection{Simulation of powder X-ray diffraction patterns and pair distribution functions}
 \label{SI:PDF_PXRD} 
  
  Powder X-ray diffraction (PXRD) patterns and pair distribution function (PDF) data were simulated using v0.9 of the \texttt{matador} package \cite{matador}. All PXRD patterns were calculated assuming a Cu-$K_\alpha$ source with wavelength 1.541\,\r{A}. PXRD peak intensities were corrected with atomic scattering factors, Lorentz-polarization correction and a thermal broadening term with an element-independent Debye-Waller factor ($B=1$). All peaks were then artificially broadened with a Lorentzian envelope of width 0.03\,$\degree$, which ignores instrument-dependent and momentum-dependent peak broadening. PDFs were computed from pairwise atomic distances in non-diagonal supercells to ensure all appropriate distances were considered within the region shown (1--8\,\r{A}). The pairwise distances were then collected as a histogram and broadened with a Gaussian envelope of width 0.01\,\r{A}. In these calculations, no element-dependence was introduced into the scattering behavior of each atom and thus intensities are not expected to match experimental data.
  
  \subsection{Details of band structure calculations}
  
  To calculate the spin-orbit coupled electronic band structures for Cu$_2$P, Ir$_2$P and Rh$_2$P, $J$-dependent pseudopotentials were used to account for coupling between orbitals, as implemented in CASTEP v20.1. All band structures were calculated using paths generated with SeeK-path \cite{HINUMA2017140}, with a $k$-point spacing of 2$\pi$ $\times$ 0.03 \AA$^{-1}$. The Cu $J$-dependent pseudopotential required a higher plane wave kinetic energy cutoff of 1000\,eV, while both Rh and Ir required only 500\,eV.

   \begin{figure}
      \centering
      \subfloat[$P6_3cm$--Cu$_3$P]{{\includegraphics[width=0.6\textwidth]{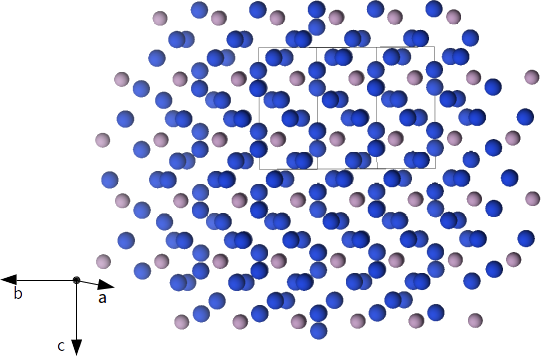}}}%

      \subfloat[$I\bar{4}3d$--Cu$_3$P]{{\includegraphics[width=0.6\textwidth]{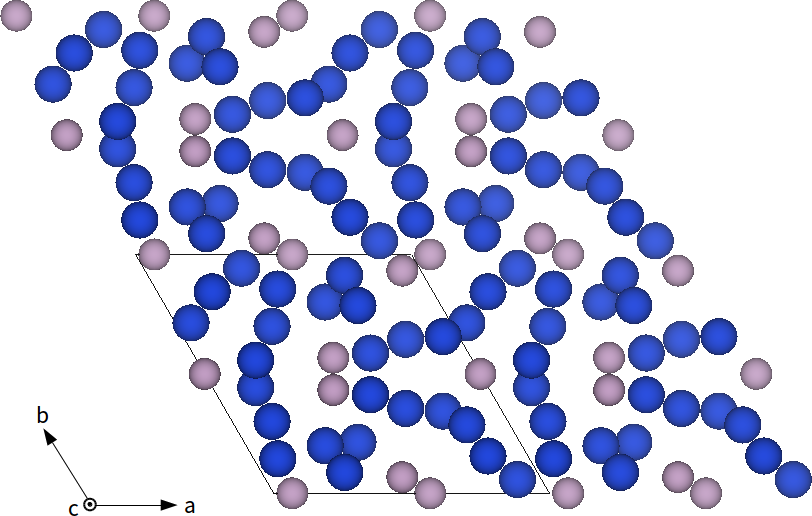} }}%
      \caption{$P6_3cm$--Cu$_3$P (ICSD 15056) and $I\bar{4}3d$--Cu$_3$P from a swap with Cu$_3$As (ICSD 64715) shown as $2\times2\times2$ supercells without Cu--P connectivity to illustrate the symmetry of both structures. }%
      \label{fig:Cu3PCu3AsICSDstructs}%
  \end{figure}
  
    \begin{figure}
    \centering
    \subfloat[]{\includegraphics[scale=1]{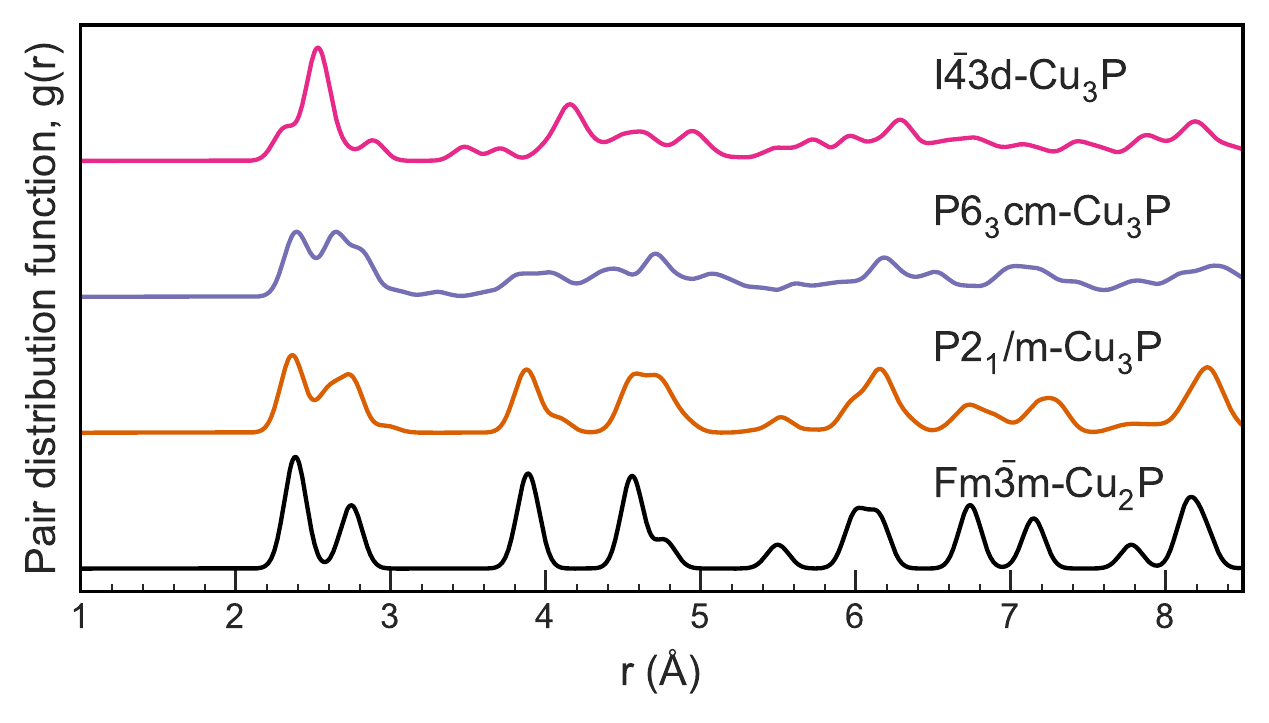}}
    
    \subfloat[]{\includegraphics[scale=1]{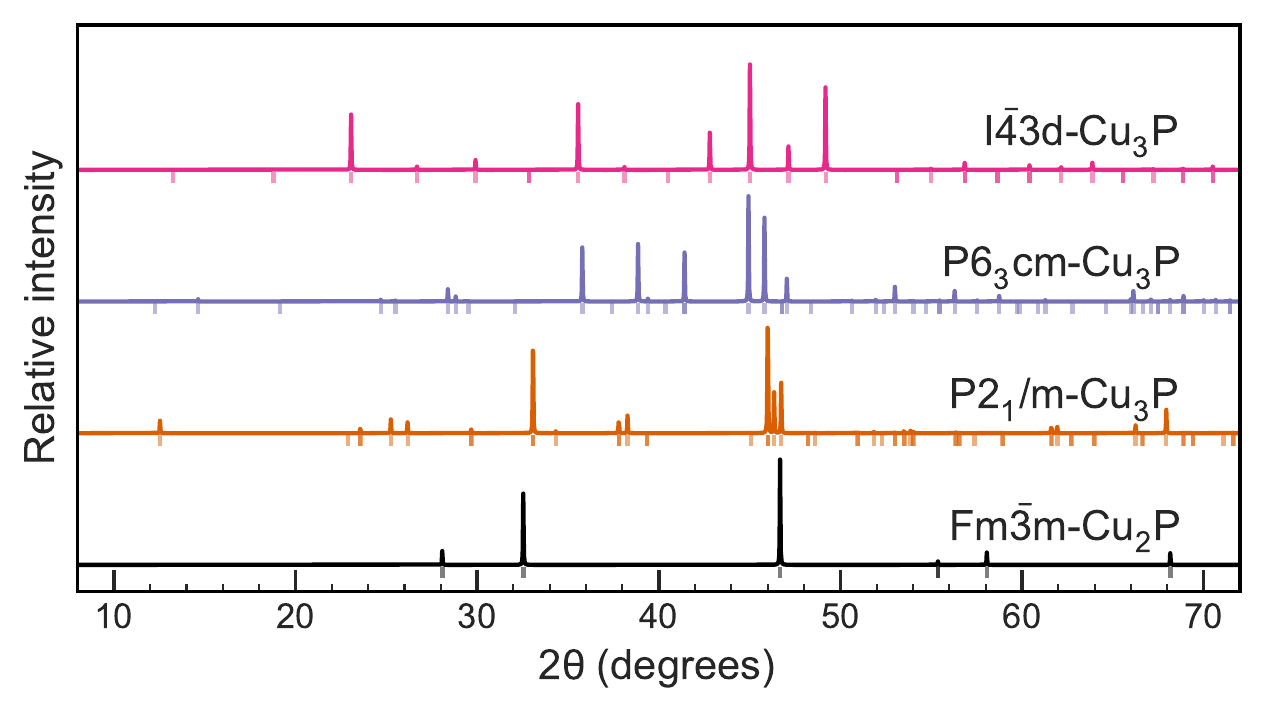}}
    
    \caption{(a) PDFs of Cu$_3$P and Cu$_2$P all have two peaks between 2 and 3\,\AA\,except $I\bar{4}3d$--Cu$_3$P which has one sharp peak at \textasciitilde2.6\,\AA . (b) PXRD patterns show peaks within 2$\degree$ peaks in either $I\bar{4}3d$ or $P6_3cm$--Cu$_3$P. Details of the PDF and PXRD calculations can be found in the Supplementary Information.}
    \label{fig:Cu2PCu3P-pdf-pxrd}
  \end{figure}
  
  \begin{figure}
    \centering
    \subfloat[]{{\includegraphics[scale=1]{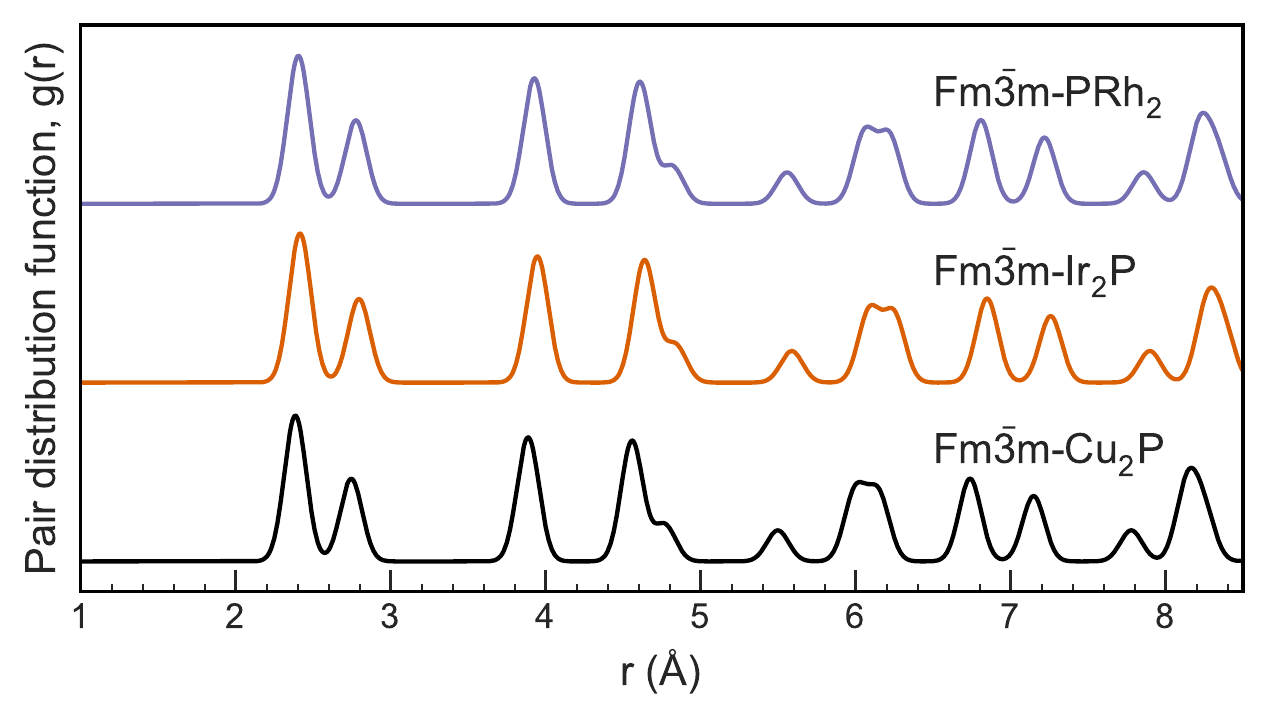}}}%
    
    \subfloat[]{{\includegraphics[scale=1]{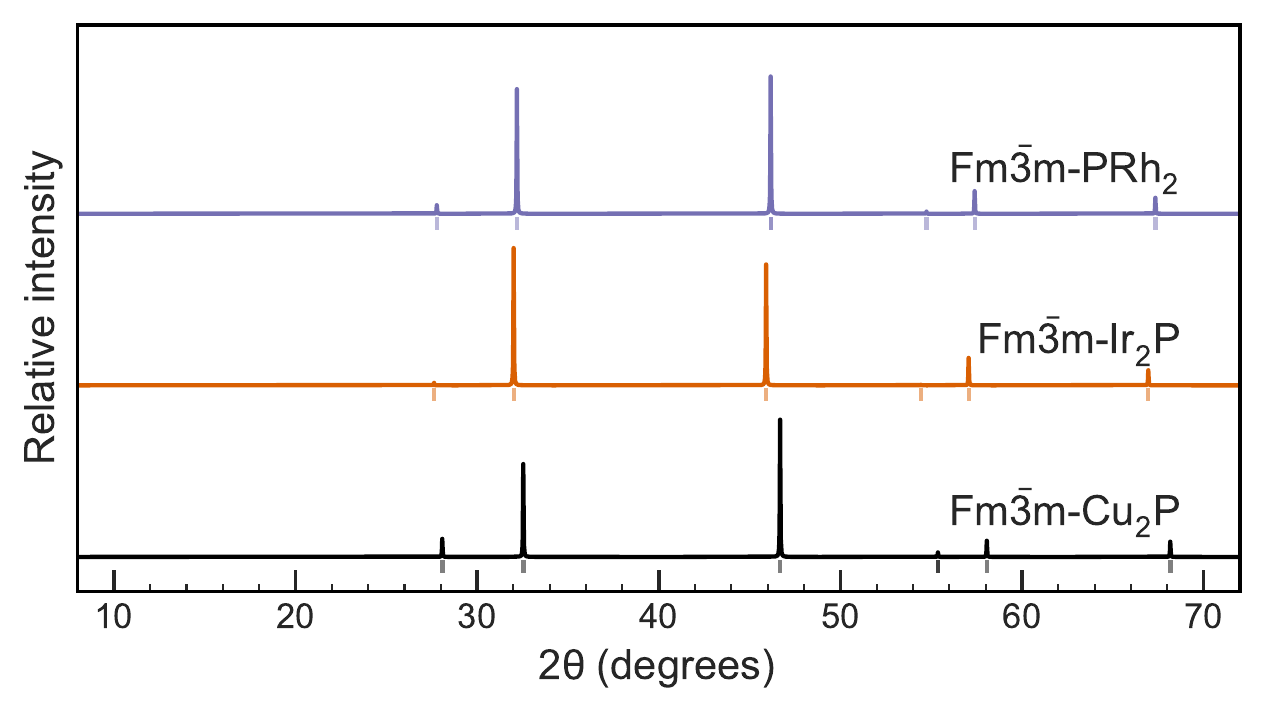} }}%
    \caption{(a) PDF and (b) PXRD plots of Cu$_2$P, Ir$_2$P, and Rh$_2$P show all three structures are nearly identical. Cu$_2$P was identified from the prototype $Fm\bar{3}m$-Rh$_2$P and Ir$_2$P structures shown here. The Cu$_2$P peak shifts to higher 2$\theta$ values are a result of a cell-shrinking during geometry optimization to a side length of 3.89\,\AA, whereas Ir$_2$P and Rh$_2$P have cell side lengths of 3.95\,\AA\;and 3.93\,\AA\;respectively. All PDFs are artificially broadened with Gaussians of width 0.1\,\AA\;and PXRDs are calculated using a Cu K$_{\alpha}$ source.}
    \label{fig:Cu2PIr2PRh2P-pdf-pxrd}
  \end{figure}
  
  \begin{figure}
    \centering
    \includegraphics[scale=1]{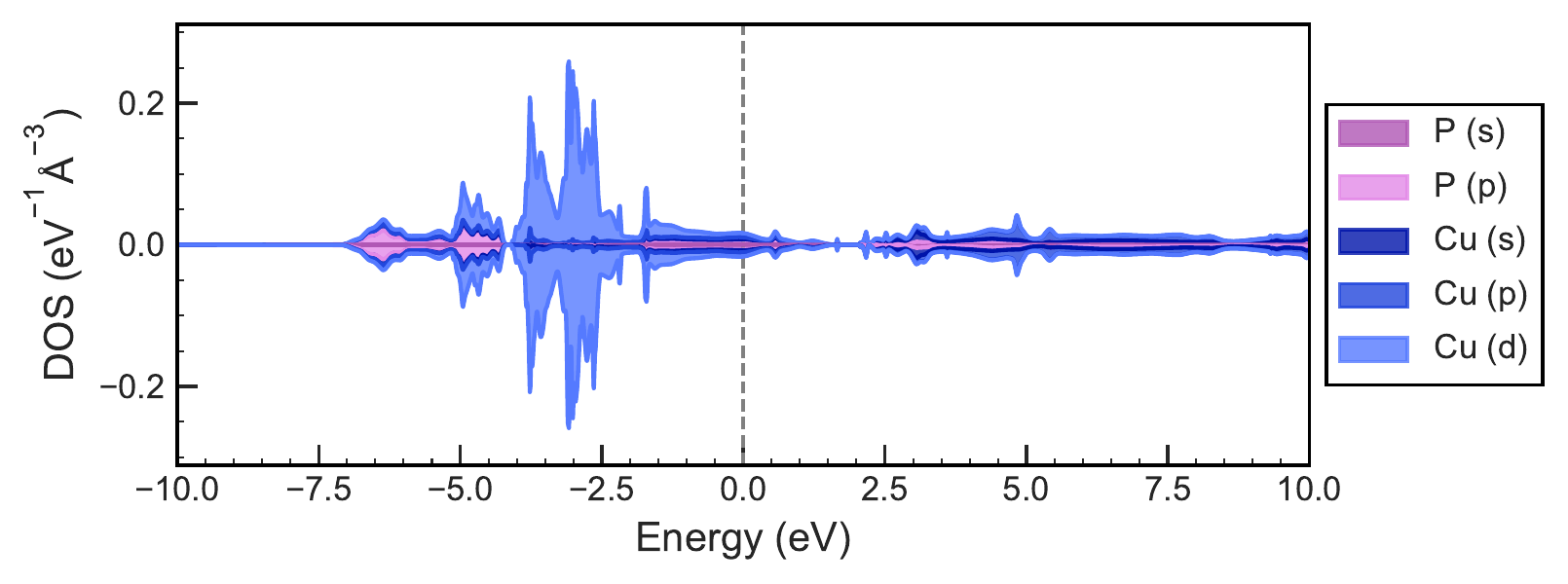}
    \caption{Cu$_2$P density of states calculated with PBE and plotted by spin channel, showing a non-magnetic state. The density of states are calculated by OptaDOS and projected onto the P $s$ and $p$ and Cu $s$, $p$, $d$ states, as well as the up and down spins.}
    \label{fig:Cu2P-spin-dos}
  \end{figure}
  
  \begin{figure}
    \centering
    \includegraphics[scale=1]{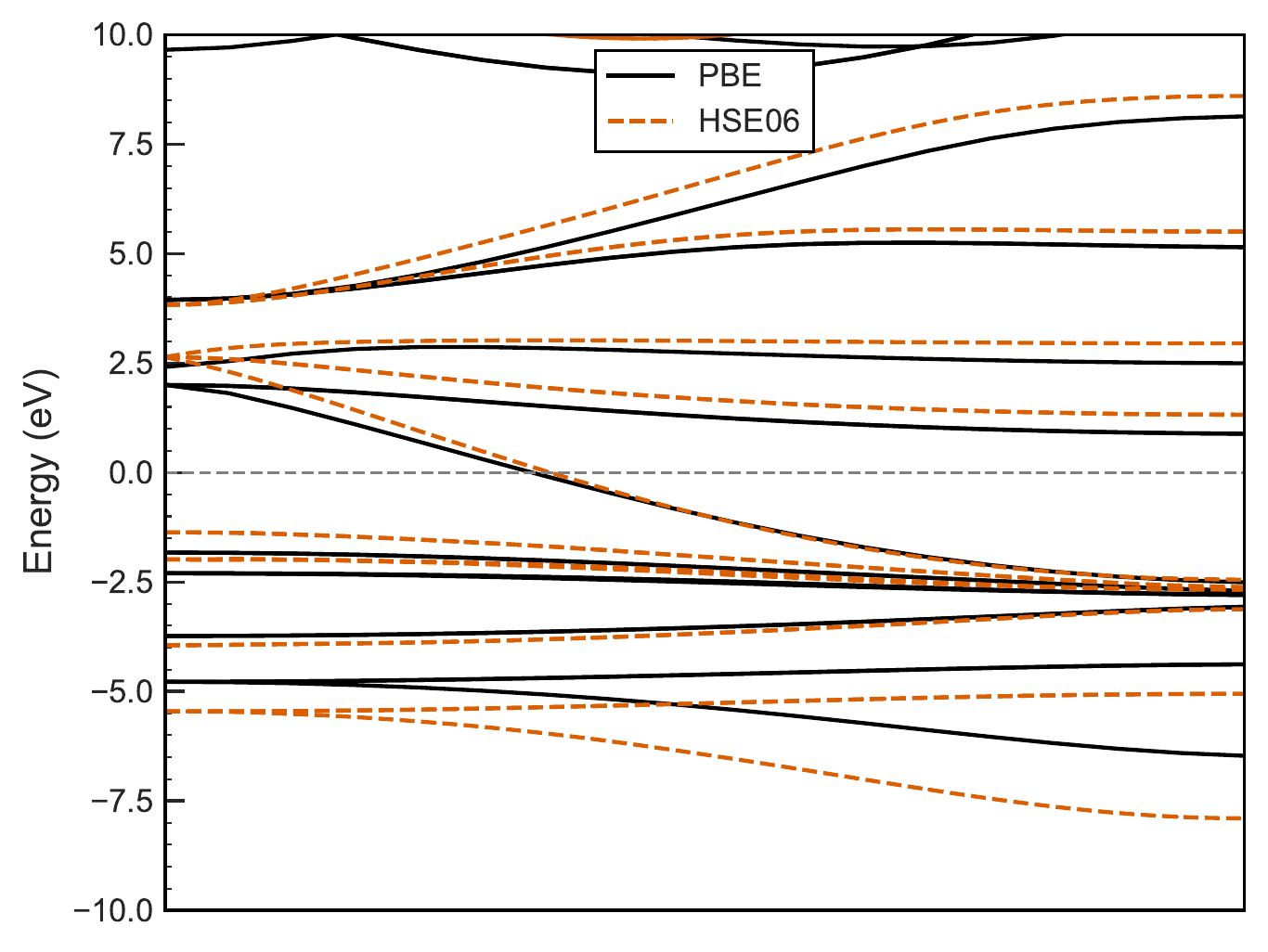}
    \caption{Comaparison between PBE and HSE06 band structures for Cu$_2$P calculated with CASTEP v18 \cite{clark2005first}. The HSE06 bands were offset by 0.35\,eV and are shown with the dashed orange line, alongside the PBE bands in black, for the  $\Gamma$ to $T$ path through the BZ. The bands are touching at 2.5\,eV at the $\Gamma$ point using the HSE06 functional where the PBE bands are non-degenerate at this point.}
    \label{fig:cu2p-hse06-bands}
  \end{figure}

   \begin{figure}
    \centering
    \subfloat[Cu$_2$P]{{\includegraphics[scale=1]{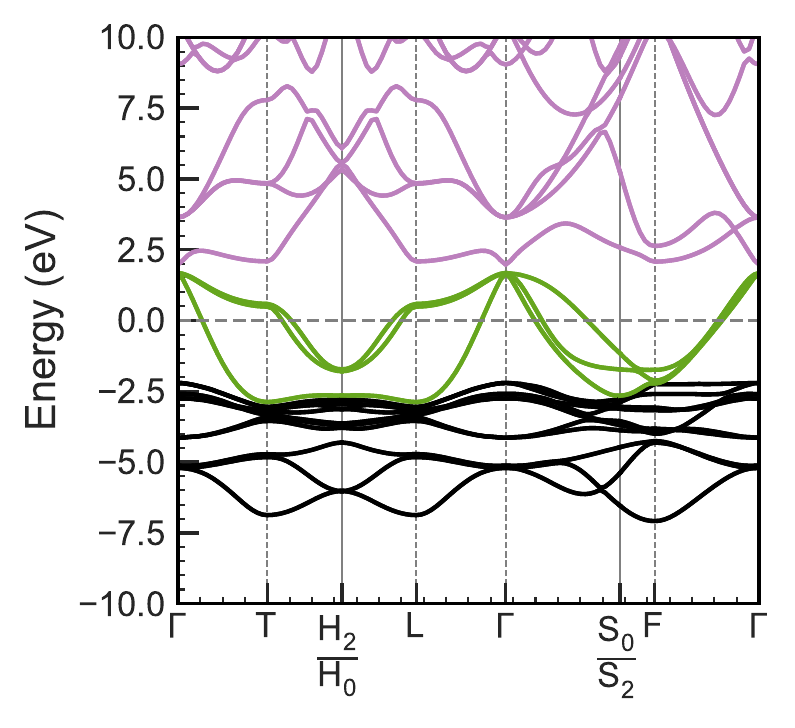}}}%
    \qquad
    \subfloat[Ir$_2$P]{{\includegraphics[scale=1]{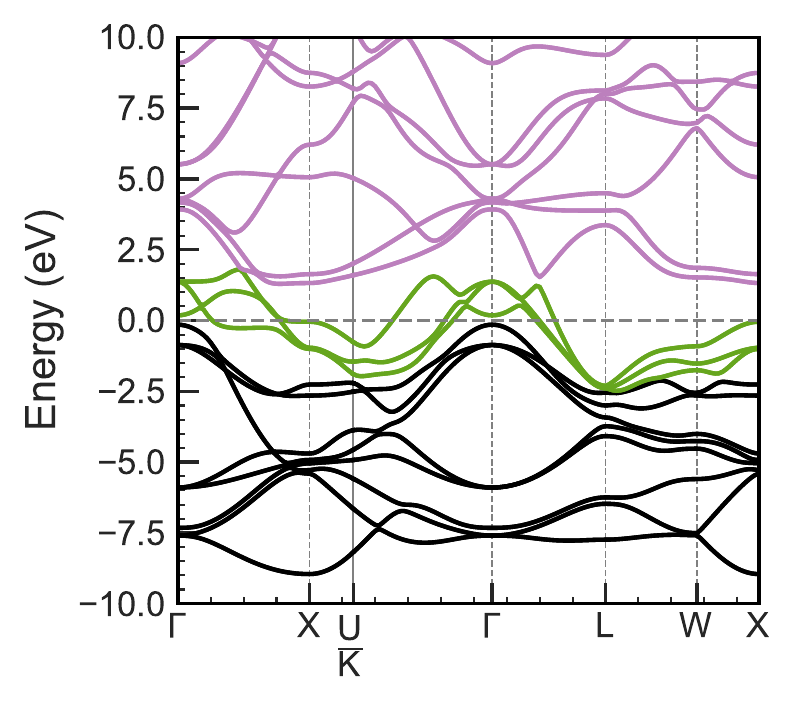}}}%
    \subfloat[Rh$_2$P]{{\includegraphics[scale=1]{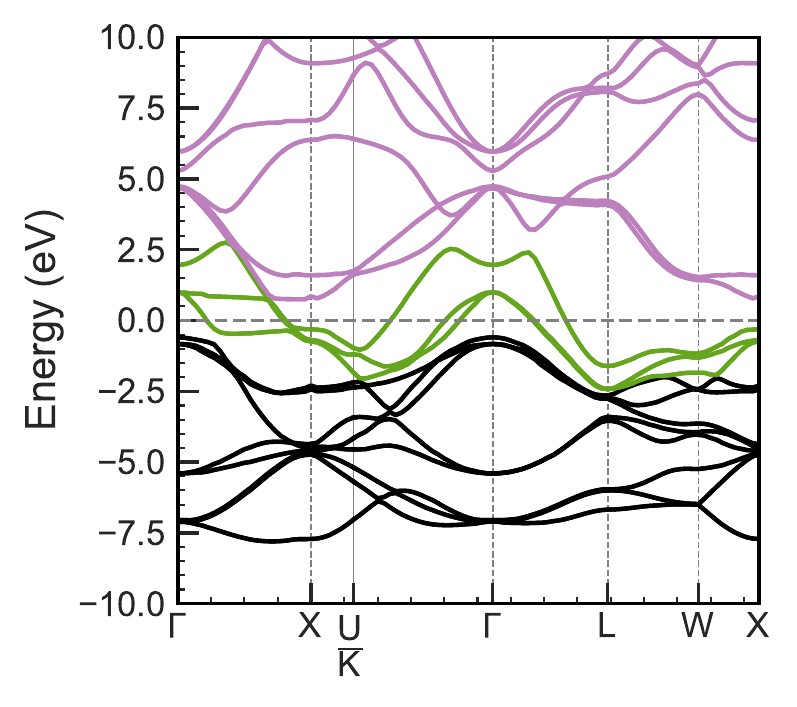} }}%
    
    \caption{Cu$_2$P, Ir$_2$P and Rh$_2$P spin-orbit coupled electronic band structures calculated with CASTEP and OptaDOS \cite{clark2005first,NichollsOptaDOS,morris2014optados}. All structures are metals with dispersive bands across the Fermi level (0.0\,eV). The black dashed box in panels (b) and (c) shows the band gap crossing between two bands at the $\Gamma$ point in Ir$_2$P and gap at the same location in Rh$_2$P.}
    \label{fig:SOC}
  \end{figure}

  \begin{figure}
    \centering
    \includegraphics[scale=1]{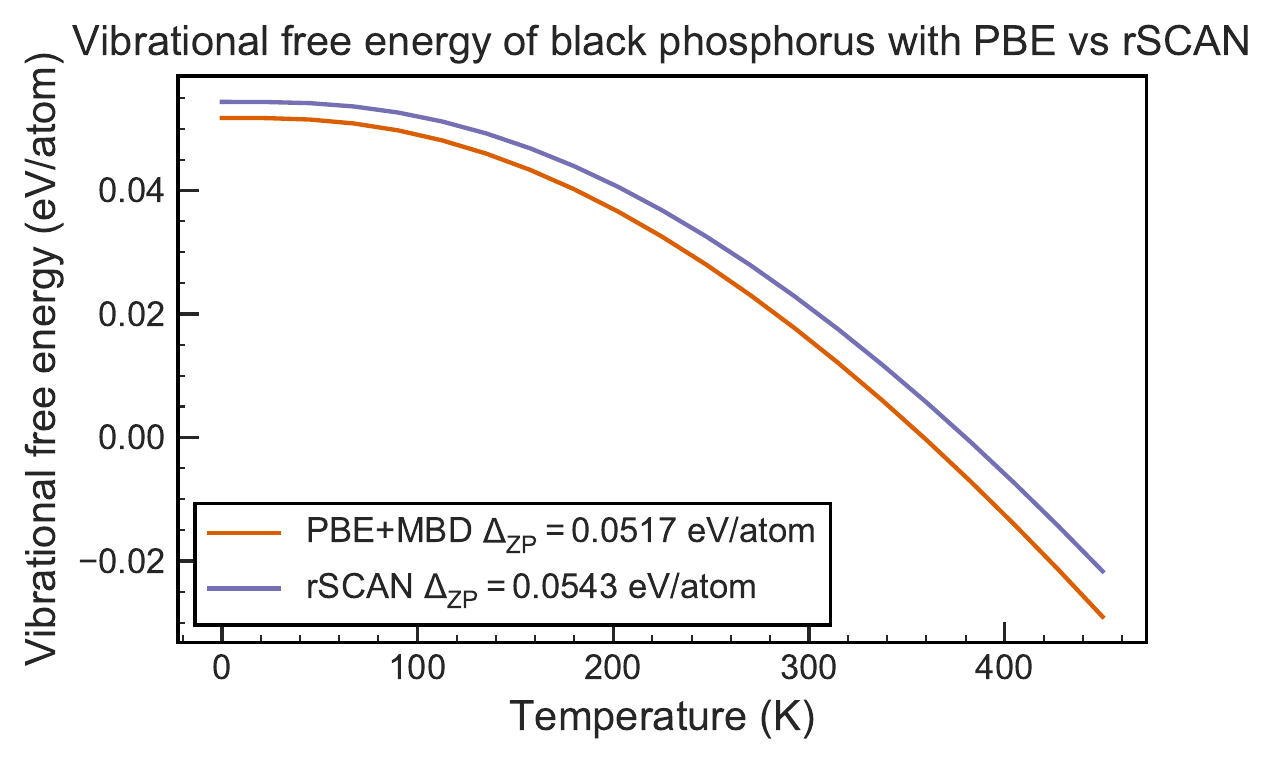}
    \caption{Comparison of free energies calculated using the rSCAN functional versus the PBE functional with MBD* correction which shows that the zero-point energy ($\Delta_{ZP}$) is the only difference, justifying the use of PBE ground-state energies with MBD* correction for the chemical potential of black phosphorus as an equivalent method to using the rSCAN functional.}
    \label{fig:scan-pbe-compare}
  \end{figure}

  \begin{figure}
    \centering
    \includegraphics[scale=1]{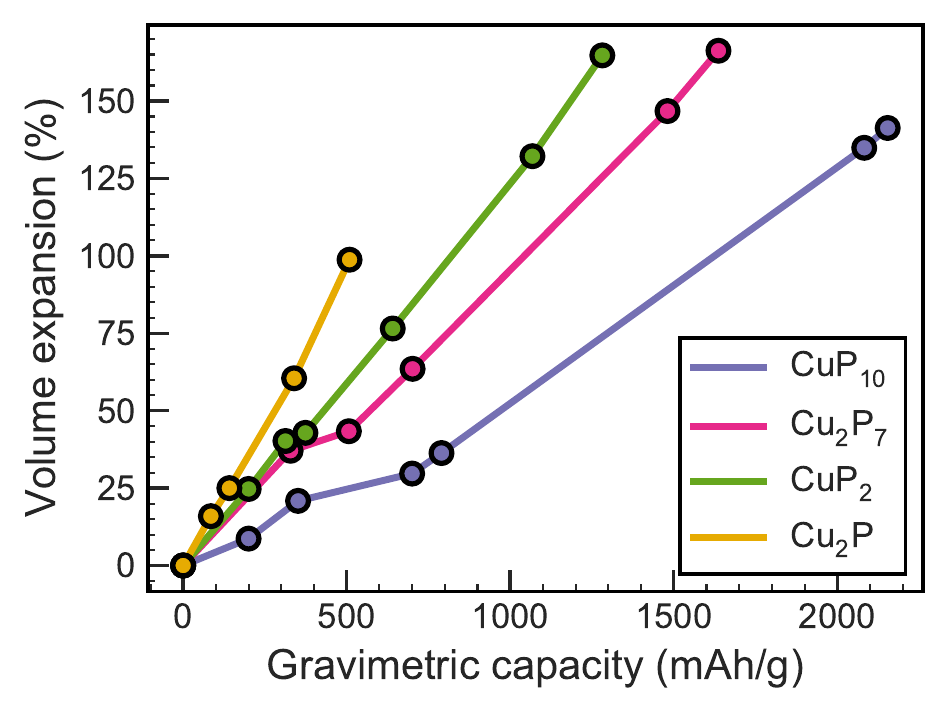}
    \caption{Volume expansion for 4 Cu--P phases. These lines follow the red pathways shown in Figure \ref{fig:LiCuP-hull} where each point represents a two-phase region of the ternary hull, as a red pathway crosses over a black tie-line.}
    \label{fig:LiCuP-volume}
  \end{figure}
  
  \begin{figure}
      \centering
      \includegraphics[scale=1]{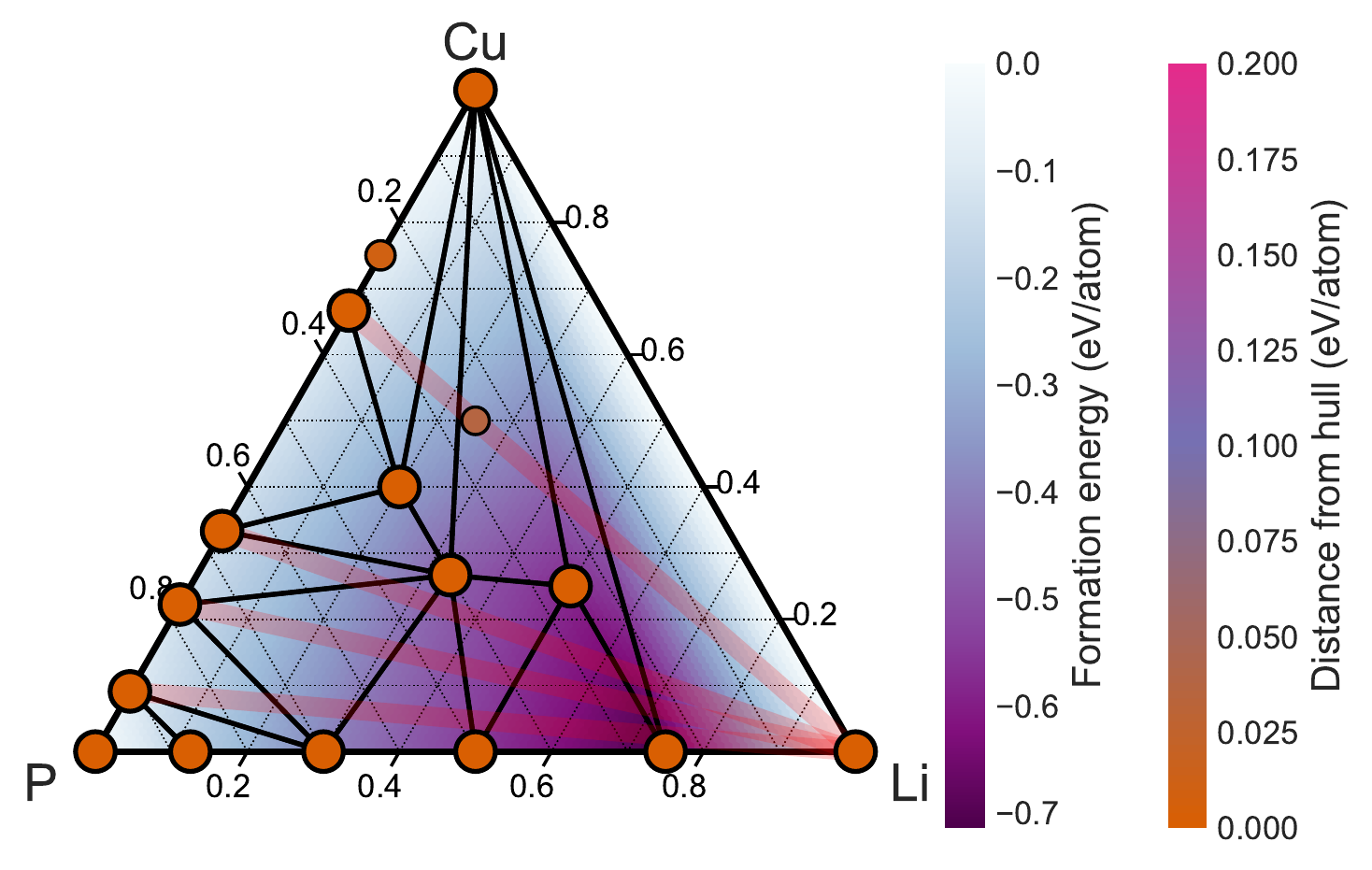}
      \caption{Ternary convex hull used in constructing the voltage profile for Cu$_2$P and Cu$_3$P. Structures on the hull are colored orange, and connected by tie-lines in black; red pathways show the ground-state conversion pathway from binary Cu--P structures to pure Li. Individual structures are represented by points which are colored by their distance from the hull. The ternary space itself is colored by the depth of the convex hull at that point. The red pathways are drawn from stable Cu--P phases towards Li; for example, this shows the phases which Cu$_2$P goes through during cycling before the final state of (Li$_3$P + 2 Cu).}
      \label{fig:LiCuP-hull}
  \end{figure}

\end{document}